\documentclass[conference]{IEEEtran}
\IEEEoverridecommandlockouts

\usepackage{amsmath,amssymb,amsfonts}
\usepackage{graphicx}
\usepackage{textcomp}
\usepackage{xcolor}
\def\BibTeX{{\rm B\kern-.05em{\sc i\kern-.025em b}\kern-.08em
		T\kern-.1667em\lower.7ex\hbox{E}\kern-.125emX}}

\usepackage{enumerate}
\usepackage{algorithm, algorithmic}
\usepackage{subfigure}

\newtheorem{definition}{Definition}

\usepackage{tikz}
\newcommand*{\circled}[1]{\lower.7ex\hbox{\tikz\draw (0pt, 0pt)%
		circle (.5em) node {\makebox[1em][c]{\small #1}};}}

\begin{document}

\title{Lauca: Generating Application-Oriented Synthetic Workloads}

\author{\IEEEauthorblockN{
		Yuming Li$^1$, Rong Zhang$^1$, Yuchen Li$^2$, Ke Shu$^3$, Shuyan Zhang$^1$, Aoying Zhou$^1$ \\
		$^1$East China Normal University, $^2$Singapore Management University, $^3$PingCAP Ltd.}
	\IEEEauthorblockA{\{liyuming@stu, rzhang@dase, syzhang@stu, ayzhou@dase\}.ecnu.edu.cn, yuchenli@smu.edu.sg, shuke@pingcap.com}
}

\maketitle

\begin{abstract}
	
	The synthetic workload is essential and critical to the performance evaluation of database systems.
	When evaluating the database performance for a specific application, the similarity between synthetic workload and real application workload determines the credibility of evaluation results.
	\iffalse
	% assessing
	Currently, the workload for performance evaluation is generally from standard benchmarks or simulated according to the actual application workload.
	The benchmark workload is an abstraction of a class of applications, and usually very different from the target application workload, so the evaluation result is not meaningful and referable.
	% informative 
	And the simulated workload cannot have the same workload characteristics as the real application due to the primitive nowaday workload generation technologies, leading the unbounded error of evaluation results.
	On the other hand, owing to data privacy issues, the online workload is also difficult to apply in the practical performance evaluation.
	\fi
	% 用下面这一句说明替换掉了上面被注释的一段
	However, the workload currently used for performance evaluation is difficult to have the same workload characteristics as the target application, which leads to inaccurate evaluation results.
	% leading the evaluation result is not meaningful and referable.
	To address this problem, we propose a work\textbf{\underline{l}}o\textbf{\underline{a}}d d\textbf{\underline{u}}pli\textbf{\underline{ca}}tor (\textbf{Lauca}) that can generate synthetic workloads with highly similar performance metrics for specific applications.
	% , realizing application-oriented database performance evaluation.
	To the best of our knowledge, Lauca is the first application-oriented transactional workload generator.
	By carefully studying the application-oriented synthetic workload generation problem, we present the key workload characteristics (transaction logic and data access distribution) of online transaction processing (OLTP) applications, and propose novel workload characterization and generation algorithms, which guarantee the high fidelity of synthetic workloads.
	We conduct extensive experiments using workloads from TPC-C, SmallBank and micro benchmarks on both MySQL and PostgreSQL databases, and experimental results show that Lauca consistently generates high-quality synthetic workloads.
	
	% With the input of transaction templates and workload traces, {\em Lauca} firstly obtains the obscure transaction logics and dynamic data access distribution automatically.
	% Then, {\em Lauca} generates the corresponding (or extended) test workload upon the database system, expecting to gain the same performance metrics as the real workload.
	% In our experiments, the evaluation results of synthetic workloads generated by {\em Lauca} are highly similar to 'real' workloads from standard benchmarks and micro benchmarks for various database systems deployed on a single node or a distributed environment.
	
\end{abstract}


\vspace{0.5mm}
\section{Introduction} \label{sec:Introduction}
\vspace{0.5mm}

% \vspace{-5mm}
\begin{figure*}
	\centering
	\includegraphics[width=6.7in]{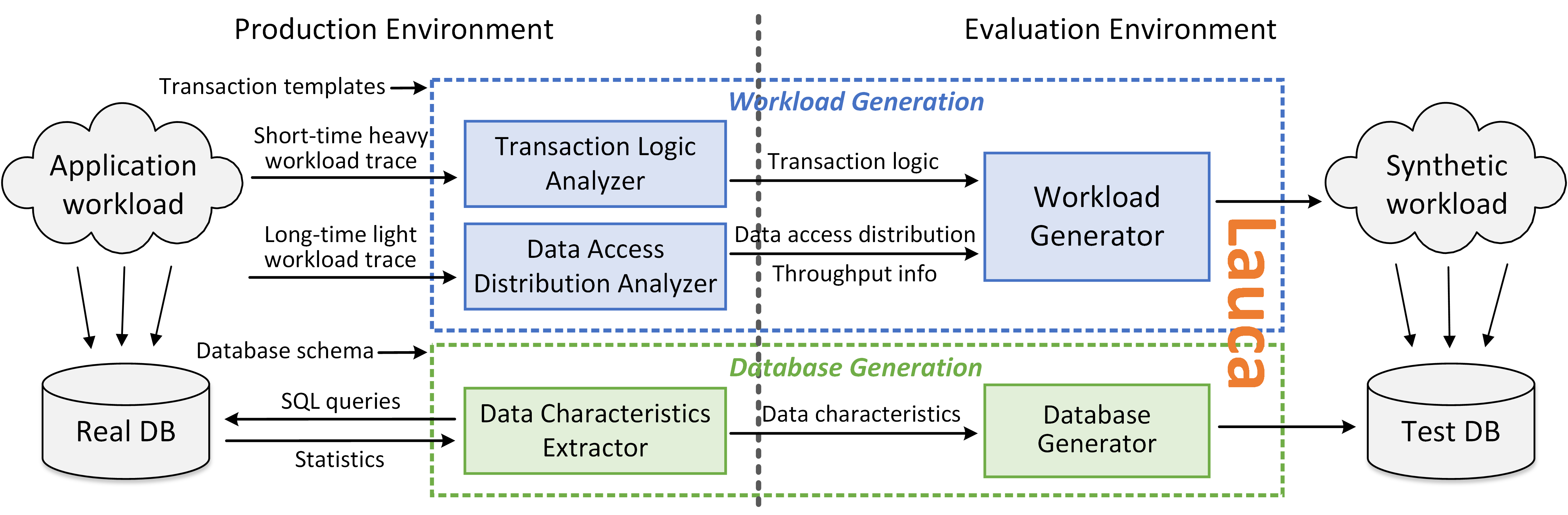}
	%\vspace{-2mm}
	\caption{The overall architecture and workflow of Lauca}
	\label{fig:lauca-arch}
	%\vspace{-2mm}
\end{figure*}

Performance evaluation is essential to the development of database management systems (DBMSs).
% Whether for database system developers or database application developers, the performance evaluation results are the important basis for us to analyze problems and make choices.
% Whether for database system developers or database application developers
For both system and application developers, performance evaluation results are the important information for analyzing problems and making choices.
% either or: 不是 ... 就是 ...; wether or: 不管 ... 还是 ...
% In view of different evaluation purposes, database performance evaluation has different ways and contents.
In this paper, we study the problem of generating synthetic workloads for application-oriented database performance evaluation, which refers to the evaluation of database system's capability for a \emph{specific} application~\cite{seltzer1999case}.
% For example, to measure the capability of a database system for a class of applications (e.g. OLTP), we can adopt standard benchmarks for relevant application area.
% Application-oriented database performance evaluation refers to the evaluation of database system's capability for a \emph{specific} application~\cite{seltzer1999case}.
The application scenarios include the followings:

% \noindent
\textbf{Database selection.} 
DBMS is the basic support system for applications, hence the application developer needs to select an appropriate system with satisfactory performance for their target workloads.
It is an effective means to evaluate candidate DBMSs by using test workloads constructed in accordance with the actual application workloads.

\textbf{PoC testing.} 
% PoC (proof of concept) generally is a realization of a certain idea to demonstrate its feasibility.
PoC (proof of concept) testing is indispensable for database vendors, especially some nascent database vendors, when marketing their products to application companies.
Building test workloads that are as similar as possible to the counterpart's applications is the most powerful guarantee for PoC conclusions.

\textbf{Application-oriented performance optimization.} 
% Generating highly realistic synthetic workload is of great significance for application-oriented database performance optimization.
Generally we can find that for application companies, database performance may not be as good as the expectation for online workloads; but for database vendors, developers cannot reproduce the performance problem during optimization because of lack of ideal evaluation workloads.

At present, there are three main ways to evaluate database performance, which are launched by using real online workloads, standard benchmark workloads and synthetic workloads, respectively.
% The evaluation workloads are real online workloads, standard benchmark workloads and synthetic workloads, respectively.
However, existing approaches are suboptimal for application-oriented database performance evaluation due to the following deficiencies:

\textbf{Data privacy issue}. 
Due to privacy concerns, it is often not possible for database vendors to obtain the real online workloads from their clients to do performance evaluation.
In fact, for the company's own testers, data privacy protection is also cumbersome.
Therefore, the evaluations have to be done using simulation workloads.

\textbf{Application oblivious}. 
One can employ standard benchmarks to evaluate database performance without data privacy issues.
Nevertheless, standard benchmarks are designed for general-purpose evaluations and their workloads are too general to imitate a specific application, and thus produce inaccurate evaluation results.

There have been some recent works on generating application-oriented synthetic workloads for online analytical processing (OLAP) applications~\cite{lo2014mybenchmark, li2018touchstone}.
However, it is still an unexplored area against OLTP applications, where the gap motivates this work.

In this paper, we propose Lauca, a transactional workload generator for application-oriented database performance evaluation.
The essential goal is that synthetic workloads generated by Lauca are as similar as possible to real application workloads, so that performance metrics evaluated by Lauca are more informative.
In order to achieve this goal, we define the key workload characteristics of OLTP applications, i.e., {\em transaction logic} and {\em data access distribution}, which determine the critical runtime behaviors of OLTP applications on database systems, including transaction conflict intensity, deadlock possibility, distributed transaction ratio and cache hit ratio.
Transaction logic is first time proposed for catching the potential business logic for fine-grained workload simulation; data access distribution is described from the perspectives of skewness, dynamics and continuity to promise the reality of synthetic workloads.
% exquisite

% 当事务是以交互式方式执行时，整个事务的所有相关业务逻辑是嵌在业务代码中的，测试人员难以从代码中完全正确地理清业务逻辑。

The workload execution model of real-world applications is also a concern for Lauca.
At present, for a considerable proportion of applications, their transaction execution is interactive, where SQL requests are sent one by one; at the same time, some of the transactions are executed using stored procedures~\cite{pavlo2017we}.
Therefore, Lauca must be able to support these two types of transaction execution models.
For this reason, we propose the concept of transaction template to allow users to more easily specify the transactions of target applications.

The overall architecture and workflow of Lauca are presented in Figure~\ref{fig:lauca-arch}.
We give an overview of Lauca from two aspects, which are synthetic database generation and synthetic workload generation.
The inputs of {\em Database Generator} are database schema and data characteristics, among which data characteristics are extracted from the real database automatically by {\em Data Characteristics Extractor} using simple SQL queries.
% Since the data characteristics are tedious and need to be obtained from the real database, we provide a {\em Data Characteristics Extractor} which can help us get this information automatically by using simple SQL queries.
The inputs of {\em Workload Generator} are transaction templates and workload statistics, where transaction templates are the sketches of transactions in the target application and workload statistics mainly include transaction logic and data access distribution.
By analyzing workload traces, we can extract the transaction logic and data access distribution for target application workloads.
The input of {\em Transaction Logic Analyzer} is short-time (e.g., ten minutes) heavy workload traces, which contain all the parameters and return items of each SQL operation.
The input of {\em Data Access Distribution Analyzer} is long-time (e.g., a full workload cycle) light workload traces, which contain only pivotal SQL parameters.
To address the data privacy issue, Lauca is designed to isolate {\em Production Environment} and {\em Evaluation Environment}, as shown in Figure~\ref{fig:lauca-arch}, where components involving real application data are only executed in the production environment by data owners.

To the best of our knowledge, Lauca is the first transactional workload generator for application-oriented database performance evaluation.
Extensive experimental results show that transaction logic and data access distribution we proposed can effectively characterize the workloads of OLTP applications, and the deviations in performance metrics between the synthetic workloads generated by Lauca and the real application workloads are consistently less than 10\%.

The rest of the paper is organized as follows:
Section~\ref{sec:preliminaries} provides the preliminaries including problem definition and key workload characteristics considered in this work.
% of application-oriented synthetic workload generation and critical workload characteristics we need to focus on during the workload generation.
Section~\ref{sec:databasegeneration} briefly introduces the database generation.
Section~\ref{sec:transactionlogic} gives the formal definition and extraction algorithm of transaction logic.
Section~\ref{sec:dataaccessdistribution} presents three types of data access distribution in Lauca.
Section~\ref{sec:workloadgeneration} shows the concrete implementation of workload generation.
Section~\ref{sec:experiments} reports the experimental results of Lauca.
Section~\ref{sec:relatedwork} reviews the related studies in the literature.
Finally, Section~\ref{sec:discussion} gives some discussions and Section~\ref{sec:conclusion} concludes the paper.

\vspace{0.5mm}
\section{Preliminaries} \label{sec:preliminaries}
\vspace{0.5mm}
%\vspace{+1mm}

% In this section, we first give the formal problem definition of application-oriented synthetic workload generation, then present the derivation process of critical workload characteristics that need control during the workload generation, and finally describe the overall architecture and workflow of {\em Lauca}.

\subsection{Problem Definition}

Before giving the formal problem definition, we enumerate the natural requirements of application-oriented database performance evaluation as follows.

\textbf{Fidelity.}
The evaluation workloads should be highly similar to the real application workloads.
\textbf{\em Performance metrics} (e.g., {\em throughput}, {\em latency}, {\em utilizations of various physical resources}, etc) obtained by the evaluation are expected to be the same as running in the real application environment.
The \textbf{\em similarity} between evaluation workloads and real application workloads is measured by the {\em deviations in performance metrics}.
The smaller the deviation, the higher the similarity.

% \textbf{Sheltering.}
% {\color{blue}{\textbf{Confidentiality.}}}
\textbf{Security.}
Data privacy protection is the basic requirement for commercial applications, therefore real data and workloads generally cannot be directly used for database performance testing.
% especially for database vendors.

% However, in order to generate synthetic workloads similar to the target applications, it is also necessary to provide some basic workload information, such as database schema and transaction templates (p.s., item names can be renamed).

\textbf{Scalability.}
The target application may have huge data scale and high request concurrency/throughput.
It requires that the workload generation toolset is scalable to multiple nodes and can support parallel database and workload generation.

% \textbf{Extending.}
\textbf{Extensibility.}
In some cases, the request concurrency and throughput of current real application workload cannot meet the test requirements, so the real application workload needs to be able to be extended by our tool for generating synthetic workloads with desired scales.

% Sometimes, we need to extend the current application workload to measure the database performance under synthetic workloads with the expected scale.
% Since this paper is aimed at transactional workloads, the main extending directions are request concurrency and request throughput.

\iffalse
\textbf{Usability.}
% performance evaluation process
The entire evaluation process should be highly automated with no manual involvement.
And the toolset should be friendly to users, even non-professionals, therefore the input of the tools should be easy to build or easy to obtain.
\fi

Based on these requirements, we formulate the problem of application-oriented synthetic workload generation in Definition~\ref{defn:problemdefi}.

\begin{definition} \label{defn:problemdefi}
	\textbf{Application-oriented synthetic workload generation:} 
	Generating synthetic workloads that are highly similar to the target application workloads, which is not only required to have small deviations in performance metrics on the databases, but also to promise the properties of security, scalability and extensibility.
\end{definition}

% This is a high-level problem definition, but we can see that the problem involves miscellaneous application workloads, complex database function implementations, numerous performance metrics, and parallel processing in distributed environments.
% As the first solution to this problem, this paper may not be perfect in theory, but provides a solution that can meet the practical evaluation requirements and most of the application scenarios.

% {\color{red}{We can see that the problem involves miscellaneous application workloads, complex database function implementations, numerous performance metrics, and parallel processing in distributed environments, which make it a hard problem to find an optimal solution.}}

\begin{figure}
	\centering
	\includegraphics[width=0.88\columnwidth]{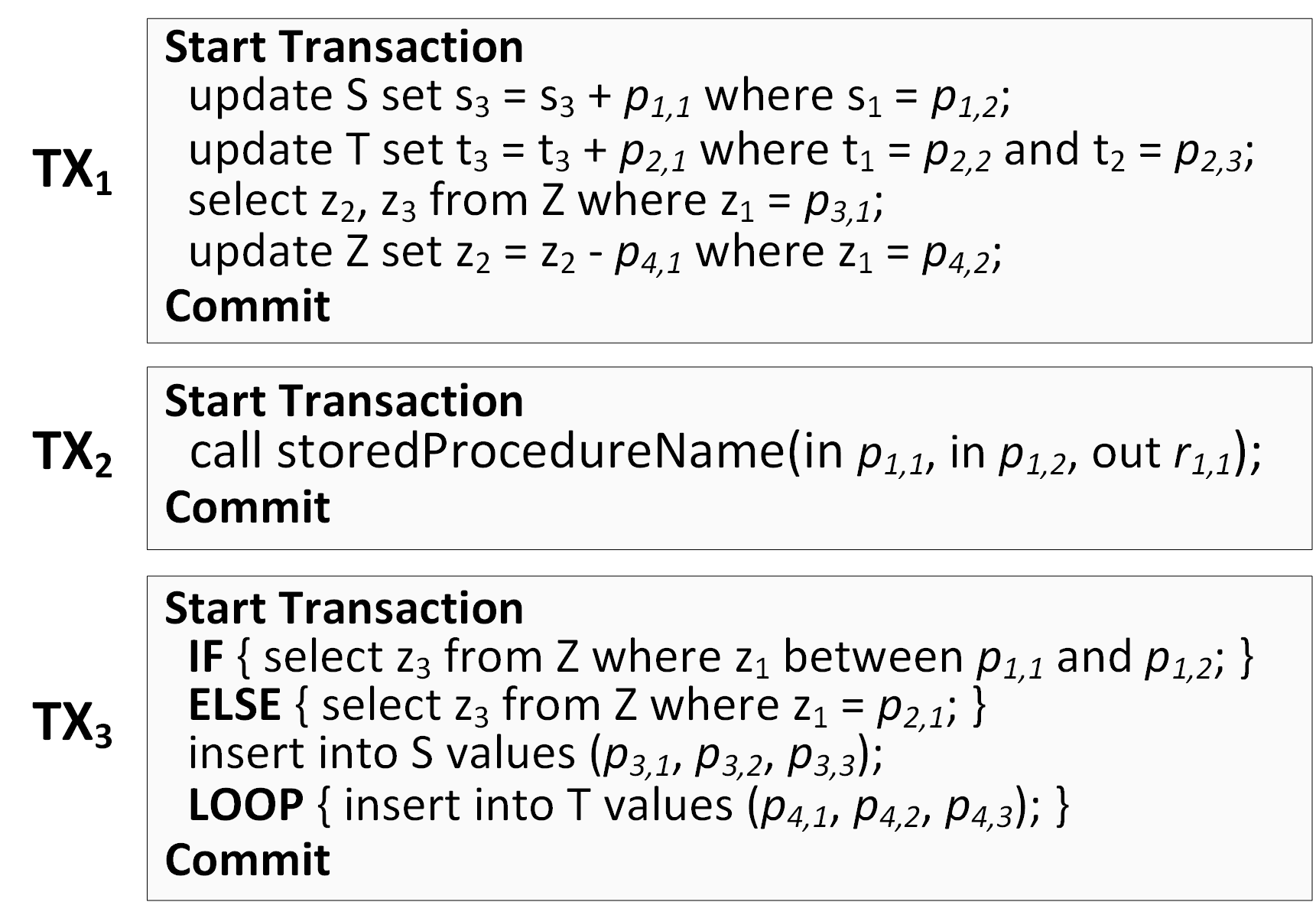}
	%\vspace{-1mm}
	\caption{Examples of transaction templates}
	%\vspace{-1mm}
	\label{fig:txtemplates}
\end{figure}

We propose transaction templates that make it easy and simple to specify the transactions of target applications.
Transaction template supports two mainstream workload execution models, i.e., interactive and stored procedure.
When transactions are executed interactively, SQL operations are sent to the database one by one, and all relevant business logics for the transactions are embedded in application codes.
Stored procedures can consolidate and centralize the logics that were originally implemented in application codes, and all applications call the procedures.
According to the survey conducted by Pavlo~\cite{pavlo2017we}, more than half of the actual applications do not use stored procedures.
This brings a challenge.
It is difficult for testers to extract precise transaction definitions like the stored procedures from long and complex application codes if stored procedures are not used.
To solve this problem, the transaction template is proposed, which allows users to define the sketches of transactions in target application without specifying exact relationships among SQL operations.
% , and transaction templates also support stored procedures well
The transaction template is defined as follows.

\iffalse
We propose the concept of transaction template to allow users to simply specify the transactions of target applications.
% The transaction template we proposed is used to easily specify the transactions of target applications.
It is difficult to extract accurate transaction definitions like the stored procedures from an expatiatory program, if stored procedures are not used for applications.
% ask the users of Lauca to
Since a considerable proportion of transactions in practical applications are not executed as stored procedures~\cite{b28SIGMOD2017}, this paper proposes the transaction template to define the skeleton diagram of these transactions without specifying exact relationships among SQL operations.
% and there are compatibility issues among stored procedures in different database systems,
The transaction template is defined as follows.
\fi

\begin{definition} \label{defn:txtemplatedefi}
	\textbf{Transaction template:} It is a transaction sketch, which consists of SQL operations and branch/loop structures (if any), where the parameters in SQL operations are symbolized.
	The SQL operation can be a stored procedure call.
	All relationships among SQL operations, parameters and return items, as well as the judgment conditions in branch and loop structures are ignored.
\end{definition}

Figure~\ref{fig:txtemplates} presents three sample transaction templates.
TX$_1$ and TX$_3$ are executed in an interactive manner, while TX$_2$ is executed as a stored procedure.
The parameters in all transaction templates are symbolized.
TX$_1$ contains 4 SQL operations, and TX$_3$ contains branch and loop structures.
If the transaction is executed as a stored procedure, the corresponding transaction template contains only one SQL operation, that is, a stored procedure call, such as TX$_2$.
In addition, the stored procedure needs to be created in the test database to support evaluation.

\subsection{Workload Characteristics}

In order to make the synthetic workloads and the real application workloads have the same execution cost on the same database system and thus have the same performance metrics, we must analyze which workload characteristics need our attention and control during synthetic workload generation.
Notice that our work is to simulate workloads for a specific application upon a certain database system. So some factors related to system performance are fixed, such as database schema, transaction templates, the implementation mechanisms of DBMS, etc.
% The \textbf{\em transaction template} consists of several SQL operations with symbol parameters, and sometimes contains conditional branches and loop structures.
We generate the synthetic workloads by instantiating the symbolic parameters in transaction templates.

% All of our play space is to generate the synthetic workload by instantiating the symbol parameters in the transaction templates and generate the test database according to the database schema and data characteristics.

\begin{figure}
	\centering
	\includegraphics[width=0.8\columnwidth]{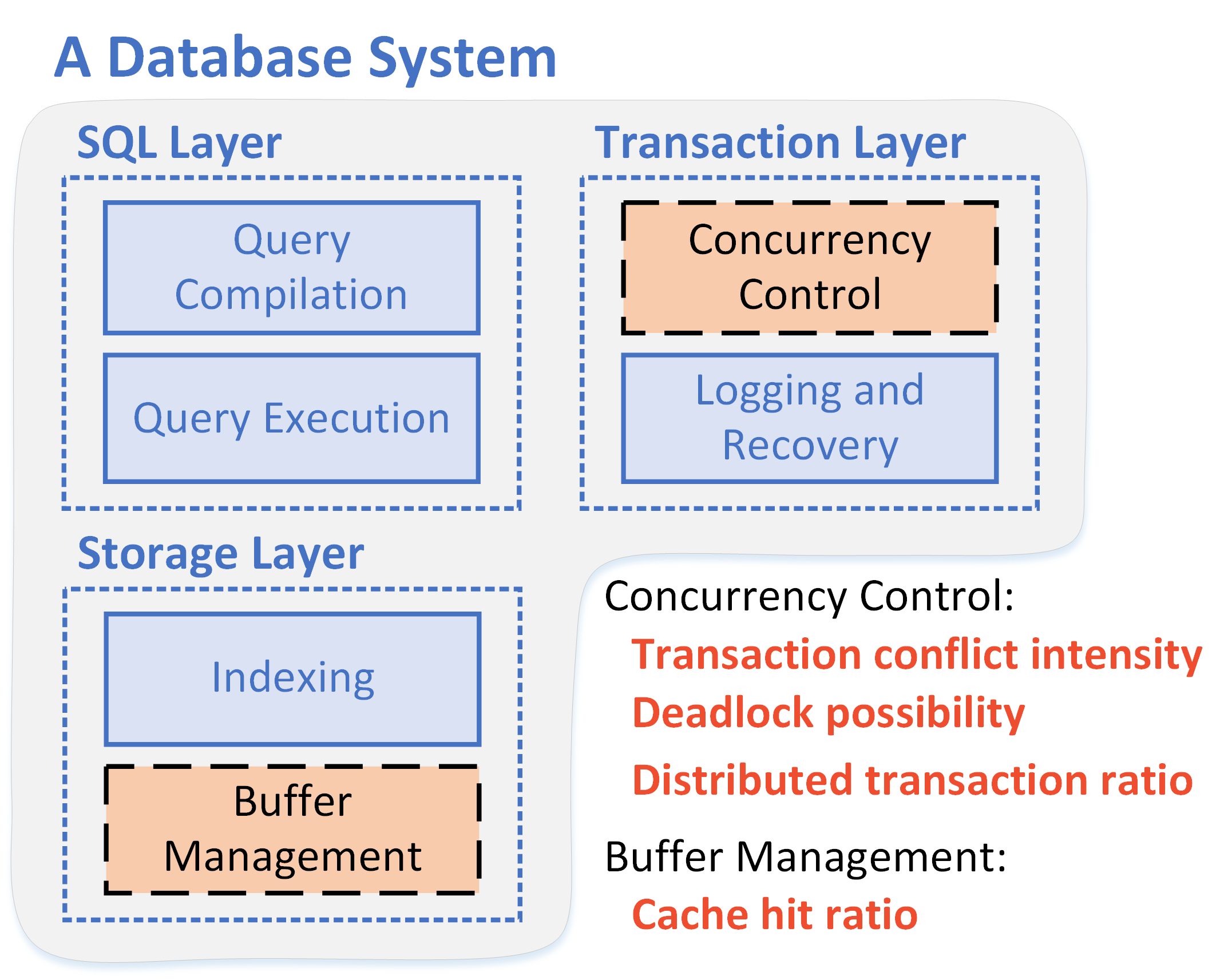}
	%\vspace{-2mm}
	\caption{Database-side workload characteristics that may be affected by different generation strategies}
	%\vspace{-3mm}
	\label{fig:db-side-wlc}
\end{figure}

A database system is mainly divided into three parts as in Figure~\ref{fig:db-side-wlc}, i.e., {\em SQL Layer}, {\em Transaction Layer} and {\em Storage Layer}.
Since we focus on transactional workloads which commonly involve no query optimizations, the generation strategies for synthetic workloads have little effect on the execution cost of SQL layer.
In the transaction layer, the workload generation strategy has a significant impact on the execution cost of {\em Concurrency Control} module.
For a stand-alone database system, the main workload characteristics that may be affected here are transaction conflict intensity and deadlock possibility.
If it is a distributed database system, we also need to pay attention to the impact on distributed transaction ratios.
For a given database system, transaction log volume is the main performance factor to {\em Logging and Recovery} module.
And the log volume is determined by transaction templates and data characteristics, which are fixed for a target application.
{\em Indexing} and {\em Buffer Management} are two main modules in the storage layer.
When database schema and transaction templates are fixed, the execution cost of indexing is settled.
However cache hit ratio in buffer management module is heavily influenced by workload generation strategies, which shall be considered as a key workload characteristic as well. % taken seriously.

\iffalse
\begin{table}
	% \footnotesize
	\centering
	\caption{Application-side workload characteristics} \label{tab:app-side-wlc}
	% \vspace{1mm}
	\small{
		\begin{tabular}{|l|l|} \hline
			\cellcolor[gray]{0.9} Database-side workload & \cellcolor[gray]{0.9} Related application-side \\
			\cellcolor[gray]{0.9} characteristics & \cellcolor[gray]{0.9} workload characteristics \\ \hline
			Transaction conflict intensity & Transaction logic, \\
			& Data access distribution \\ \hline
			Deadlock possibility & Transaction logic, \\
			& Data access distribution \\ \hline
			Distributed transaction ratio & Transaction logic \\ \hline
			Cache hit ratio & Data access distribution \\ \hline
	\end{tabular}}
\end{table}
\fi

In order to ensure that the synthetic workload and the real application workload are consistent on these four database-side workload characteristics, we define and manipulate two application-side workload characteristics, namely {\em transaction logic} and {\em data access distribution}.
% For each database-side workload characteristic, the related application-side workload characteristics are shown in Table~\ref{tab:app-side-wlc}.
Transaction logic depicts the relationship among SQL parameters and return items, and it is a representation of the potential business logic of target applications.
The potential business logic, such as data items accessed by different SQL operations in a transaction satisfy a certain correlation in a probability, usually cannot be seen at a glance and need to be obtained by analyzing workload traces.
Transaction logic has a big impact on transaction conflict intensity, deadlock possibility and distributed transaction ratio.
The technical details about transaction logic are available in Section~\ref{sec:transactionlogic}.
Data access distribution is used to characterize the access distribution of SQL operations on data items.
We analyze the workload traces to get data access distributions, namely the data distributions of SQL parameters, for parameter instantiation during the workload generation.
Data access distribution has a big impact on transaction conflict intensity, deadlock possibility and cache hit ratio.
Section~\ref{sec:dataaccessdistribution} presents all the details about data access distribution.

\vspace{1mm}
\section{Database Generation} \label{sec:databasegeneration}
\vspace{1mm}

% Currently, two usual ways can be used to obtain these data characteristics, one is to query on original data tables with simple SQL queries, e.g., 'SELECT COUNT(*) FROM table' for table size; the other is to query on the database statistics, provided that there are such statistics.

% All data characteristics needed for database generation are: for tables, there is the table size; for all types of columns, there are the percentage of Null values and the cardinality of unique values; for numeric columns, there is the domain; for string typed columns, there are the average length and the maximum length; for boolean typed columns, there are the percentages of True and False values.

All data characteristics needed for database generation are the same as used in Touchstone~\cite{li2018touchstone}, such as table size, column domain, column cardinality of unique values, etc.
The generation of test database is actually the generation of multiple tables, while satisfying primary/foreign key constraints, as well as data characteristics of non-key columns.
% There is no new contribution in this section to the previous work~\cite{li2018touchstone}.
% Covering database generation is mainly for the completeness of Lauca functionality.
Introducing database generation is mainly for the completeness of Lauca functionality and the later description of synthetic workload generation, and our database generation is similar to previous work~\cite{li2018touchstone}.

% In Lauca, the generation of each table can be divided into two parts, one is the generation of primary and foreign keys, and the other is the generation of non-key columns.
% Below we use an example to describe the generation mechanism of primary and foreign keys, and introduce how to generate the non-key columns with desired data characteristics using random column generators.

% 所有需要的数据特征与工作1一样，比如有表大小，属性阈值，非重复值个数，字符串平均长度和最大长度。

\begin{figure}
	\centering
	\includegraphics[width=0.95\columnwidth]{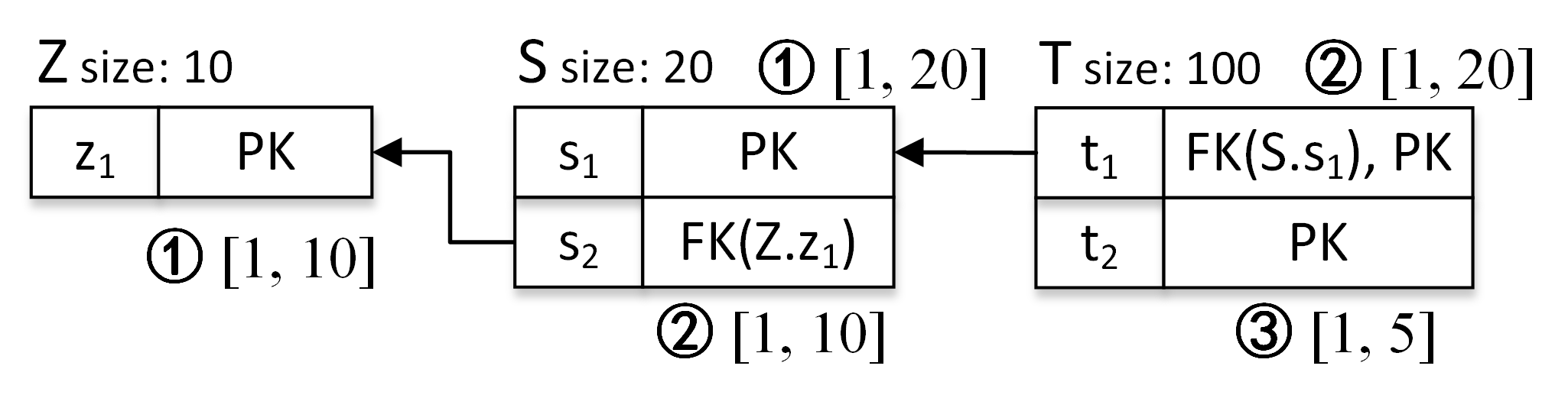}
	%\vspace{-3mm}
	\caption{Determining the domains of key columns}
	%\vspace{-3mm}
	\label{fig:key_generation}
\end{figure}

Without loss of generality, we assume that primary/foreign key columns are integers.
The primary key is the identifier of the record and generally has no physical meaning, so we do not need to consider the data characteristics of these key columns.
We simply generate primary keys sequentially and generate foreign keys within the referenced primary key domain randomly, which can promise the uniqueness of primary keys and the referential integrity of foreign keys.
Before the generation of key columns, there are three steps to determine their domains, with an example shown in Figure~\ref{fig:key_generation}.
Firstly, if the primary key contains only a single column, the domain is $[1, s]$, where $s$ is the table size (as{\circled {\small 1}}in Figure~\ref{fig:key_generation}).
Secondly, the domains of foreign key columns can be determined by the domains of referenced primary key columns (as{\circled {\small 2}}in Figure~\ref{fig:key_generation}).
Thirdly, we settle down the domains of non-foreign key columns in the composite primary key, e.g., column $T.t_2$ (as{\circled {\small 3}}in Figure~\ref{fig:key_generation}).
The most usual and reasonable situation is that there is only one non-foreign key column in the composite primary key.
And the domain of this column is $[1, \frac{s}{\prod |d_{fk}|}]$, where $d_{fk}$ is the domain of one of the foreign key columns within the composite primary key.
Other situations are handled similarly.
Due to cascaded references, the second and third steps may need to be performed multiple times.
% to the composite primary key

The random column generator~\cite{li2018touchstone}, which contains a random index generator and a value transformer, is used to generate values for non-key columns, while satisfying the desired data characteristics, especially the cardinality of unique values.
The output of random index generator is the integers from 1 to $n$, where $n$ is the column cardinality.
Given an index, the transformer deterministically maps it to a value in the domain of the column.
We adopt different transformers based on the data type of the column.
For numeric types, e.g., integer, we simply pick up a linear function which uniformly maps the index to the column domain.
For string types, e.g., varchar, there are some seed strings pregenerated randomly, which satisfy the length requirements.
We first select a seed string based on the input index, such as the $(i\%k)^{th}$ seed string, where $i$ is the index and $k$ is the number of all seed strings.
Then we concatenate the index and the selected seed string as the output value.
% This approach allows us to easily control the cardinality of string typed columns with tiny memory consumption.
% The generation of Null values and boolean typed columns are based on simple probability control.

In summary, the generation of each table is independent of each other.
And for each table, we can achieve parallel data generation with multiple threads on multiple nodes by assigning a primary key range to each thread.

% 为避免基于非主键的谓词没有匹配的记录，我们最好可以确保所有的基数索引和外键值都尽可能的出现。???

\vspace{1mm}
\section{Transaction Logic} \label{sec:transactionlogic}
\vspace{1mm}

The transaction logic proposed in this paper is not an extension of predicate logic in previous work~\cite{bonner1998logic}, but rather a representation of potential business logic for OLTP applications.
% The input transaction templates are the representation of explicit business logic.
% In this section
Here, we first give an intuitive explanation of why transaction logic matters through a concrete example; then we give a formal definition of transaction logic; finally, we present how to efficiently extract transaction logic from workload traces.
% clearly give

% 先给宏观上的思路 ->  再给明确的定义 ->  再给示例说明，关键是写清楚，为什么这么做？最后还得分析这个完备性以及可扩展性？

\subsection{Intuitive View} \label{sec:intuitiveview}

\iffalse
The following is a sample transaction template (abbr., TT) for illustrating the necessity of transaction logic.

{\small \slshape 
	{\bfseries Start Transaction}
	
	update $S$ set $s_3$ = $s_3$ + {\color{blue}{$p_{1,1}$}} where $s_1$ = {\color{blue}{$p_{1,2}$}};
	
	update $T$ set $t_3$ = $t_3$ + {\color{blue}{$p_{2,1}$}} where $t_1$ = {\color{blue}{$p_{2,2}$}} and $t_2$ = {\color{blue}{$p_{2,3}$}};
	
	select {\color{blue}{$q_2$}}, {\color{blue}{$q_3$}} from {\color{blue}{$Q$}} where {\color{blue}{$q_1$}} = {\color{blue}{$p_{3,1}$}};
	
	update {\color{blue}{$Q$}} set {\color{blue}{$q_2$}} = {\color{blue}{$q_2$}} - {\color{blue}{$p_{4,1}$}} where {\color{blue}{$q_1$}} = {\color{blue}{$p_{4,2}$}};
	
	{\bfseries Commit}
}
\fi

\begin{figure}
	\centering
	\includegraphics[width=0.88\columnwidth]{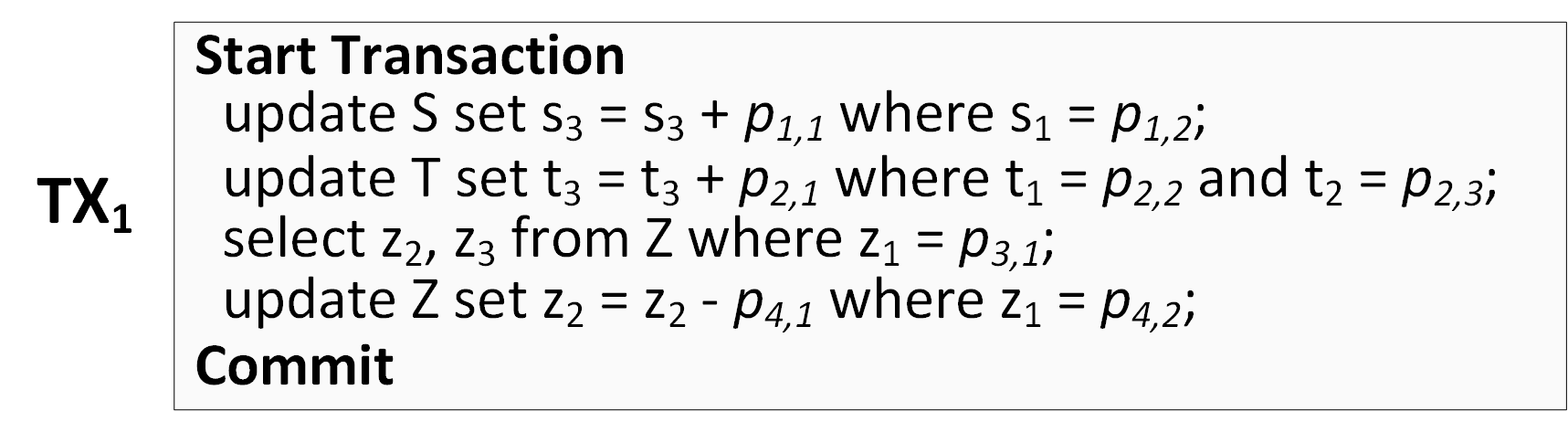}
	%\vspace{-2mm}
	\caption{Transaction template TX$_1$ from Figure~\ref{fig:txtemplates}}
	%\vspace{-3mm}
	\label{fig:tx1templates}
\end{figure}

We use the sample transaction template TX$_1$ in Figure~\ref{fig:tx1templates} to illustrate the necessity of transaction logic.
% The tables $Z$, $S$ and $T$ in TX$_1$ can refer to Figure~\ref{fig:key_generation}.
Referential relationships among tables $Z$, $S$ and $T$ in TX$_1$ are represented in Figure~\ref{fig:key_generation}.
$z_2$, $z_3$, $s_3$ and $t_3$ are double typed non-key columns.
$p_{i,j}$ is the $j^{th}$ symbolic parameter of $i^{th}$ SQL operation. % , where $i$ and $j$ are both count from 1
When generating synthetic workloads based on TX$_1$, we usually instantiate the symbolic parameters with random values generated according to standard data distributions.
% Because the primary key is continuous, there must be a record that satisfies the predicate, that is, all SQL operations must fall on a specific data item without missing.
% The parameters in all set clauses are generally set to a small random value to avoid the column being out of the domain.
% In such a way, our synthetic workload is generated.
However, there may be significant performance differences between such synthetic workloads and real application workloads.
Now suppose we have deployed a distributed database, and tables $Z$, $S$, and $T$ are all hash-partitioned by the first column of primary key.
Due to business logics, there are usually some correlations between SQL parameters in real workloads.
For example, parameters $p_{1,2}$ and $p_{2,2}$ are of the same value in a high probability (e.g., 99\%) in actually.
Thus the first two SQL operations in TX$_1$ do not bring many distributed transactions.
But in the synthetic workloads, the above two parameters are randomly generated, and they are likely to be different, which will lead to serious distributed transactions.
As we know, the ratio of distributed transactions has a crucial impact on database performance~\cite{harding2017evaluation}.
More than that, if parameters $p_{3,1}$ and $p_{4,2}$ in the last two SQL operations of TX$_1$ are always the same in the real workload, there will be no deadlocks.
However, these two parameters may be different during the synthetic workload generation, which can lead to deadlocks.% resulting in deadlocks possibly.
% (assume that the concurrency control protocol is 2PL and the isolation level is serializable).
The occurrence of deadlocks can seriously affect the database performance~\cite{krivokapic1999deadlock}.
Through this example, it is intuitive that the potential business logic of the target application, namely the transaction logic we proposed, is an important workload characteristic for synthetic workload generation, and must be guaranteed to be consistent with the real application workload.

%\vspace{-1mm}
\subsection{Definition of Transaction Logic} \label{sec:txlogicdefinition}

The workloads of actual applications vary greatly, and the potential business logic is even more difficult to depict.
Therefore, it calls for a general framework to model transaction logic that abstracts from different applications.
% the definition of transaction logic must jump out of the perspective of application business.
We propose to define the transaction logic using {\em transaction structure information} and {\em parameter dependency information}.
Branches and loops are common structures in transactions.
% Their particular conditional judgments are not important for synthetic workload generation.
% We only care about the probability that each branch is executed and the average times that loop operations are executed.
The probability that each branch will be executed and the average numbers that loop operations will be executed have important influence on transaction execution cost. % overhead
These need to be described in the transaction structure information.

The relationship among SQL parameters and return items in a transaction describes the hidden semantics among/in SQL operations.
Even if the transaction is executed as a stored procedure, the relationship among parameters of the stored procedure call may still affect the execution cost of the transaction.
There are four types of relationships we care about after the investigation of existing OLTP benchmark workloads and actual application workloads around.
Firstly, `\textbf {equal relationship}' is the most common one.
For example, two SQL parameters are equal under a certain probability.
Secondly, `{\bfseries inclusive relationship}' is also familiar.
Because the SQL result set may be a set of tuples, the value of a SQL parameter might be an element in returned values of one previous return item.
Thirdly, `{\bfseries linear relationship}' refers to the relationship between two items, e.g., two parameters, that can be represented by a linear function, and it is a complement and extension to the equal relationship. %, which has stronger expressive ability.
Fourthly, `{\bfseries between relationship}' is proposed for predicates like `$col$ $between$ $p_{i,j}$ $and$ $p_{i,j+1}$', in which $p_{i,j+1}$ has a between relationship with $p_{i,j}$.
% and `$col$ $\geq$ $p_{i,j}$ $and$ $col$ $\leq$ $p_{i,j+1}$'
In order to simplify the problem, we only consider the relationship (i.e., dependency) between current parameter and the previous parameters (or return items).

We give the formal definition of transaction logic in Definition~\ref{defn:txlogic}.
Let's start with some terminology.
$O_i$ represents the $i^{th}$ SQL operation in the transaction template;
$r_{i,j}$ (resp., $p_{i,j}$) is the $j^{th}$ return item (resp., parameter) of $O_i$; %, where $i$ and $j$ are both count from 1;
$\Delta$ denotes the average increment of one parameter relative to another;
{\small $Pr$} represents the probability that the parameter dependency should be satisfied;
$(a, b)$ are two coefficients for characterizing the linear function.
We use the abbreviations ER, IR, LR and BR to denote the equal relationship, inclusive relationship, linear relationship and between relationship, respectively.

\begin{definition} \label{defn:txlogic}
	\textbf{Transaction logic:} 
	For a transaction template, the transaction logic consists of transaction structure information and parameter dependency information, which are specified as:
\end{definition}

\begin{itemize}
	\item \underline{Transaction structure}:
	\begin{enumerate}[{\small TS}1]
		\item Execution probability of each branch in branch structures.
		\item Average number of executions for operations in loop structures.
		% \item Rollback probability of each active rollback location.
	\end{enumerate}
	
	\item \underline{Parameter dependencies} for each parameter $p_{i,j}$:
	\begin{enumerate}[{\small PD}1]
		\item \ [$p_{i,l}$, $p_{i,j}$, BR, $\Delta$]
		\item \ A list of {\em dep-item}, where {\em dep-item} $\in$ \{ \\
		\ [$p_{m,n}$, $p_{i,j}$, ER, {\small $Pr$}], [$p_{m,n}$, $p_{i,j}$, LR, {\small $Pr$}, $(a, b)$],\\
		\ [$r_{u,v}$, $p_{i,j}$, ER, {\small $Pr$}], [$r_{u,v}$, $p_{i,j}$, LR, {\small $Pr$}, $(a, b)$], \\
		\ [$r_{u,v}$, $p_{i,j}$, IR, {\small $Pr$}] \}; \\
		$u < i$; $m \leq i$; and if $m = i$, then $n < j$.
		\item \ A list of [$p_{i,j}$, LR, {\small $Pr$}, $(a, b)$], where $O_i$ must be an operation in the loop structure.
	\end{enumerate}	 
\end{itemize}

\begin{figure}
	\centering
	\includegraphics[width=0.88\columnwidth]{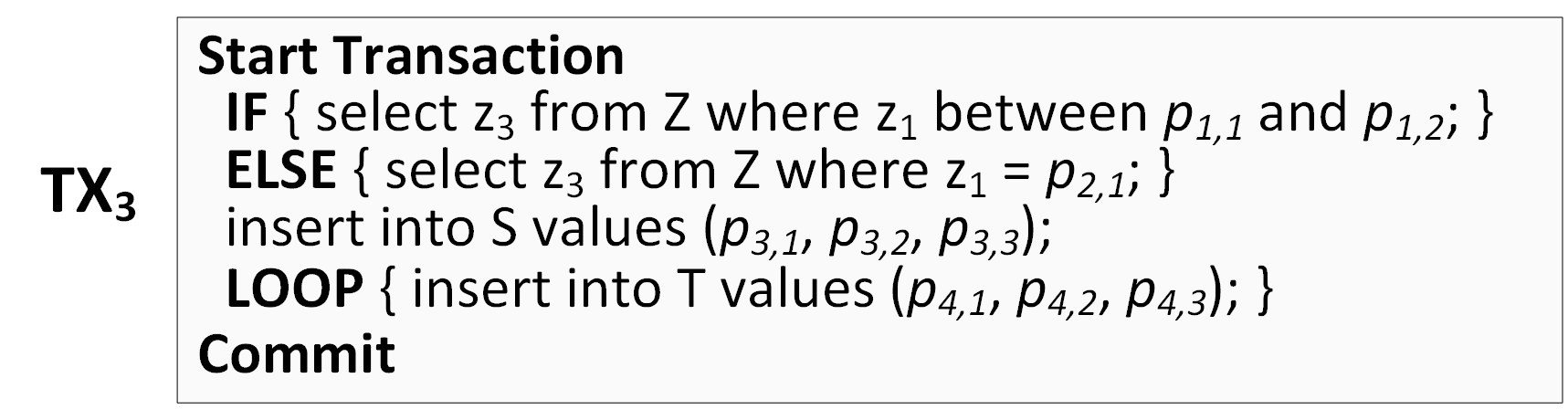}
	%\vspace{-2mm}
	\caption{Transaction template TX$_3$ from Figure~\ref{fig:txtemplates}}
	%\vspace{-2mm}
	\label{fig:tx3templates}
\end{figure}

Now, we give a more specific description to the transaction logic in Definition~\ref{defn:txlogic} using sample transaction templates.
The transaction template TX$_3$ in Figure~\ref{fig:tx3templates} contains a branch structure and a loop structure.
According to the transaction structure information defined in Definition~\ref{defn:txlogic}, we need to know the execution probability of each branch and the average number of executions of each loop structure.
This information can be obtained by analyzing workload traces.
For example in TX$_3$, the branch execution probability of the first SQL is 40\%, and that of the second SQL is 60\%; for the loop structure, the average number of executions of the forth SQL is 10.

According to the parameter dependency information defined in Definition~\ref{defn:txlogic}, there are three sets of dependencies, i.e., {\small PD}1-{\small PD}3, for each parameter in transaction templates.
{\small PD}1 is the between dependency.
If a parameter $p_{i,j}$ has {\small PD}1, then both {\small PD}2 and {\small PD}3 do not need to exist because $p_{i,j}$ can be deterministically represented by $(p_{i,l} + \Delta)$.
For example in Figure~\ref{fig:tx3templates}, $p_{1,2}$ in TX$_3$ has a between dependency [$p_{1,1}$, $p_{1,2}$, BR, 8], and the value of $p_{1,2}$ is expressed by ($p_{1,1}$ + 8).

{\small PD}2 includes equal, linear and inclusive dependencies.
For a parameter $p_{i,j}$, it may have multiple dependencies with previous parameters or return items.
Each dependency has a probability value, i.e., {\small $Pr$}, which indicates the degree of dependence.
If {\small $Pr$} is 100\%, the value of current parameter is completely determined by this dependency.
For instance in Section~\ref{sec:intuitiveview}, {\small PD}2 of parameter $p_{2,2}$ in TX$_1$ includes [$p_{1,2}$, $p_{2,2}$, ER, 99\%], and there is a dependency [$p_{3,1}$, $p_{4,2}$, ER, 100\%] for parameter $p_{4,2}$ in TX$_1$.

Moreover, since operations in loop structures are usually performed multiple times, we need to consider the value changes of SQL parameters in these operations during loop execution. % are usually delete ?
{\small PD}3 is used to represent the linear relationships between the values of the same parameter in successive runs for loop operations.
For example in TX$_3$, during the loop execution of the fourth SQL, $p_{4,1}$ remains unchanged and $p_{4,2}$ is increased by 1 each time to ensure the uniqueness of the composite primary key.
Therefore, we have {\small PD}3 [$p_{4,1}$, LR, 100\%, $(1, 0)$] and [$p_{4,2}$, LR, 100\%, $(1, 1)$] for parameters $p_{4,1}$ and $p_{4,2}$, respectively.
In addition, if a parameter that is not in an operation inside a loop structure, it can only have {\small PD}1 or {\small PD}2.

\subsection{Extraction Algorithm}

% 需要考虑事务日志可能因为回滚而不完整

% The analysis of transaction logic relies on full workload traces that contain input parameters and returned result set for each operation during the transaction execution.
The transaction logic is the representation of application business logic, so it generally does not change frequently over time. % embodiment
Therefore, the extraction of transaction logic does not require workload traces across a long time.
% We only need short-time workload traces, or the workload traces of a certain number of, e.g., 10000, transaction instances for each transaction template.
Since the transaction logic analysis of each transaction template is independent of each other, the following extraction algorithm will be introduced for a single transaction template.
Our extraction algorithm consists of six steps.
The input is a transaction template and corresponding workload traces of $K$ transaction instances.

% \vspace{+1mm}
\textbf{Step 1: Structure information.}
By traversing workload traces, we can count the number of executions for each operation in the transaction template, and use it to calculate the execution probability of each branch and the average numbers of executions for loop operations.

% \vspace{+1mm}
\textbf{Step 2: Identify BR.}
First, we identify all the parameter pairs, e.g., $<$$p_{i,l}$, $p_{i,j}$$>$, that satisfy the between relationship in transaction templates, and then traverse workload traces to get the average increment, i.e., $\Delta$, for each pair.
After that, we construct {\small DP}1 for $p_{i,j}$, and the subsequent steps 3-6 can skip the processing of $p_{i,j}$.
% 为什么仅统计average increment就可以了，tp中scan range一般很小，负载涉及的数据量大致差不多就可以了

% \vspace{+1mm}
% \noindent
\textbf{Step 3: Collect statistics for ER and IR.}
For each parameter $p_{i,j}$ in the transaction template, we traverse its previous parameters, e.g., $p_{m,n}$ (resp., return items, e.g., $r_{u,v}$), and count the number of transaction instances in which the pairs, e.g., $<$$p_{m,n}$, $p_{i,j}$$>$ (resp., $<$$r_{u,v}$, $p_{i,j}$$>$), satisfy the equal relationship (resp., equal relationship or inclusive relationship).
% [$<$$r_{2,3}$, $p_{4,1}$$>$, ER, 100] is an example statistical result indicating that there are 100 ones among the $K$ transaction instances where $p_{4,1}$ and $r_{2,3}$ are equal.
% Suppose $O_4$ is not in a branch structure, we can construct the dependency [$r_{2,3}$, ER, $\frac{100}{K}$] for $p_{4,1}$ based on the above result.
Assuming that there are totally $V$ parameters together with return items in the transaction template, there are nearly $\frac{V^2}{2}$ numbers of above pairs.
% $<$$p_{m,n}$, $p_{i,j}$$>$ and $<$$r_{x,y}$, $p_{i,j}$$>$
The complexity of step 3 is O($KV^2$).

% \vspace{+1mm}
% \noindent
\textbf{Step 4: Collect statistics for LR.}
LR only involves numeric typed parameters and return items, and the return items must be from the operations based on primary key filtering.
% now -> removed
% , and have nothing to do with the return items from the operations based on the non-primary key filtering.
Since the calculation of coefficients $(a, b)$ for LR requires two transaction instances, we randomly select $N$ groups of transaction instances (two for each group) from $K$ transaction instances.
Then we calculate the coefficients $(a, b)$ for each pair (i.e., $<$$p_{m,n}$, $p_{i,j}$$>$, $<$$r_{u,v}$, $p_{i,j}$$>$) based on $N$ groups of transaction instances.
% If there are multiple results $(a, b)$ for one pair, only the result with the most occurrences is recorded.
Notice that the LR with coefficients $(1, 0)$ needs to be ignored here, because it has been represented by ER.
% [$<$$p_{2,1}$, $p_{3,2}$$>$, LR, 100, $(1, 2)$] is an example statistical result indicating that there are 100 ones among $N$ groups of transaction instances where $p_{3,2}=p_{2,1}+2$.
The complexity of step 4 is O($NV^2$).

% \vspace{+1mm}
\textbf{Step 5: Determine ER, IR and LR by trade-offs.}
With the statistics obtained in steps 3-4, we can easily construct {\small DP}2 for each parameter $p_{i,j}$.
However, there may be a lot of dependencies for each parameter, and some of which may have very small probabilities, so we need to make a trade-off among these dependencies to remove noises and alleviate following calculations.
% (i.e., {\color{blue}{{\small $Pr$}}})
% After we abandon those dependencies with small $\xi$, we normalize the probabilities of all left dependencies for each parameter to serve the synthetic workload generation.
% In experiment section, we can find that ER is much more important than IR and LR.
We pick the most important dependencies, such as those with high probability, and ensure that the sum of probabilities of picked dependencies is less than or equal to 1.
% We are more inclined to reserve ER, and we can find that ER is much more important than IR and LR in experiment section.

\iffalse
There are four principles in the trade-off: 
1) the dependency with a large $prob$ is preferentially retained; 
2) the ER is preferentially retained relative to IR and LR due to the more significant impact on the database performance; 
3) the probability sum of all retained dependencies needs to be less than or equal to 1;
4) the retained dependencies can not be too much, e.g., up to 10.
For example, we sort all the dependencies of a parameter in descending order of $prob$, where the equal dependencies are sorted with $2*prob$.
Then, for the parameter, we select the dependencies in order until enough dependencies are included or the probability sum will exceed 1.
% 这一部分主观性过强，是否应该简化写一下，比如仅做归一化
\fi

% \vspace{+1mm}
% \noindent
\textbf{Step 6: Construct LR for loop structure.}
For multiple runs of operations in a loop structure, we use {\small DP}3 to characterize the change of parameter values.
By traversing workload traces, the value changes in successive runs are counted for each SQL parameter in the loop structure.
The calculation of coefficients $(a, b)$ is similar to the step 4.
% There may be many different variations (or increments) for a parameter, but we generally only keep some of the most frequent variations, such as the top 10.
Then based on the statistics, we construct {\small DP}3 and maintain it independently with {\small DP}2 for parameters in loop operations.
% Suppose there is an example dependency [$p_{5,3}$, LR, 90\%, $(1, 1)$], which indicates the probability that $p_{5,3}$ is increased by 1 relative to the value in last run is 90\% during the loop execution.

One thing to note is that if steps 3-4 encounter an operation in the loop structure, only the trace data of the first execution is used.
The complexity of our extraction algorithm is dominated by steps 3-4, which is O($KV^2$+$NV^2$).
In practical evaluations, $K$ and $N$ can be set to tens of thousands, and $V$ may be tens or hundreds, so the extraction of transaction logic is very fast.
In our experiments, we found that the extraction of transaction logic often takes only a few seconds with $K=10^4$ and $N=10^4$.

\section{Data Access Distribution} \label{sec:dataaccessdistribution}

Data access distribution has long been considered as an important application workload characteristic~\cite{difallah2013oltp}.
In this section, we present how to characterize and manipulate the skewness (Section~\ref{sec:s-dist}), dynamics (Section~\ref{sec:d-dist}) and continuity (Section~\ref{sec:c-dist}) of data access distribution for synthetic workload generation.

\subsection{Skewed Data Access Distribution} \label{sec:s-dist}

% As introduced in Section~\ref{sec:laucaoverview}, 
Synthetic workload generation is actually the instantiation of symbolic parameters in transaction templates.
Therefore the data access distribution of synthetic workload is determined by the values of these instantiated parameters.
Our analysis of data access distribution only needs the light workload traces containing pivotal parameters which are used to index the data involved in SQL operations.
In the following content, we present S-Dist for describing the skewness of data access, which has a serious impact on database performance.

% But not all parameters have an impact on the data access distribution, such as the varchar typed parameters corresponding to the detailed comment information.
% So our analysis of data access distribution only needs the light workload traces containing pivotal parameters, e.g., the parameters in the where clause.
% Generally, the parameters in the where clause and the parameters on the key columns are called location parameters.
% Identifying the location parameter is only to reduce the analysis cost of data access distribution.
% Even if the analysis is conducted on the all parameters, there is no any negative impact on the correctness.
% In the following content, we present the Skewed Data Access Distribution (abbreviated S-Dist) that can effectively characterize the data access skewness which has a serious impact on database performance.
% ~\cite{tian2018contention}

\begin{figure}
	\centering
	\includegraphics[width=0.82\columnwidth]{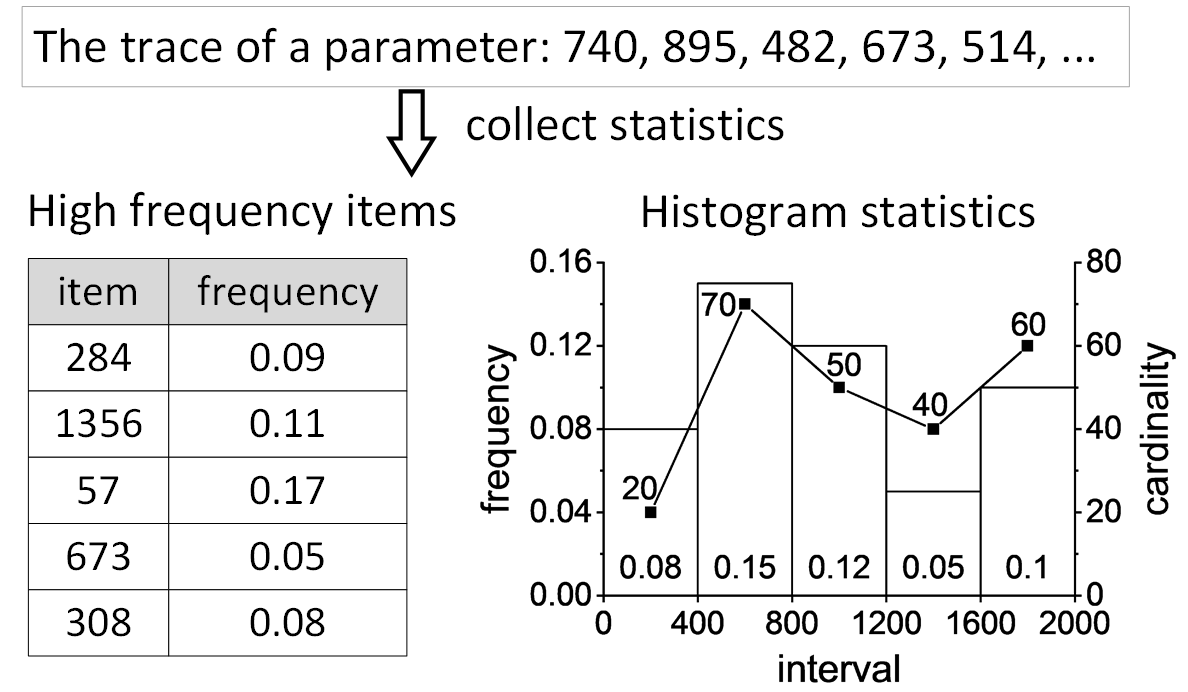}
	% \vspace{-2mm}
	\caption{An example for illustrating S-Dist}
	%\vspace{-2mm}
	\label{fig:basic_distribution}
\end{figure}

Without loss of generality, we assume the form of predicates which determine the data access distribution, for OLTP application workloads, can be expressed as `$col$ $op$ $para$'.
We use {\em high frequency items} (abbr., HFI) and {\em histogram statistics} (abbr., HS) extracted from workload traces to represent S-Dist for each parameter.
Histogram is a commonly used statistical method for density estimation, which are widely used to represent data distribution statistics in industry databases, such as Oracle, MySQL and PostgreSQL.
In S-Dist, HFI records $H$ hottest data items (i.e., concrete parameter values) with the highest occurrence frequency.
The column domain is evenly divided into $I$ intervals, and then the frequency and cardinality of the parameters, except the ones in HFI, falling on each interval are counted for HS.
Two statistics, i.e., frequency and cardinality, are used to accurately characterize the access skewness for intervals.
For instance, although the frequency of an interval is not very high, the cardinality of this interval is very small, and then parameter values in this interval can still cause high conflicts.
Figure~\ref{fig:basic_distribution} is an example S-Dist for an integer typed column with domain [0, 2000].
Both $H$ and $I$ are $5$ here.
In HFI, the hottest item is $57$ with frequency $0.17$;
in the first interval of HS, there are $20$ unique parameter values with the total access frequency $0.08$.

% We can instantiate the symbolic parameters to generate synthetic workloads using the S-Dist.
% If we use S-Dist to generate parameters directly, it will result in a large number of instantiated equality predicates (i.e., '$col$ $=$ $para$') without matching records.

% icde revision时将下面这一段简单重写了下

\begin{figure}
	\centering
	\includegraphics[width=0.98\columnwidth]{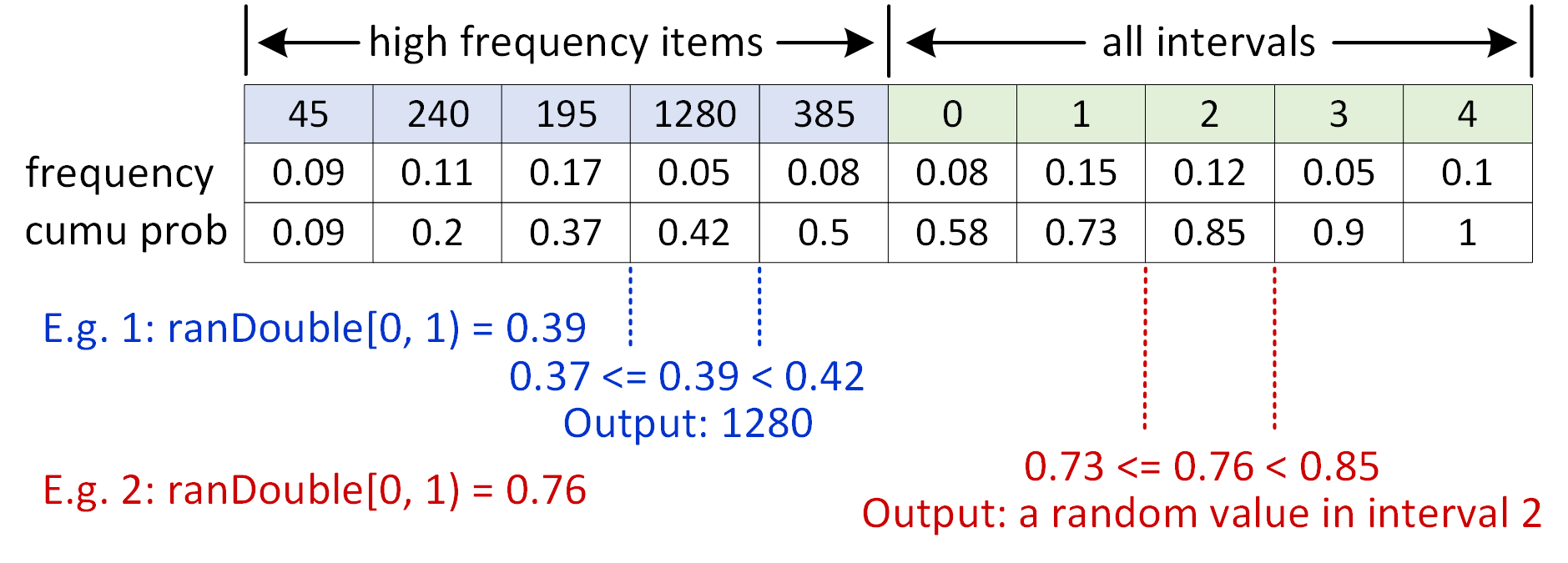}
	%\vspace{-3mm}
	\caption{An example of parameter generation}
	%\vspace{-5mm}
	\label{fig:parameter_generation}
\end{figure}

Before using S-Dist to instantiate the symbolic parameters, we need first do data transformation for HFI, because the data items in HFI are extracted from real workload traces and may not exist in generated synthetic database.
Suppose the column generator for our example is `index = ranInt[1, 400], value = index * 5', where 400 is the column cardinality, as introduced in Section~\ref{sec:databasegeneration}.
We regenerate the data items in HFI using our column generator, as shown in Figure~\ref{fig:parameter_generation}, where, for example, data item 57 is replaced by 195.
For ensuring that the generated parameters conform to the desired frequency distribution, we calculate a cumulative probability array based on the frequencies of high frequency items and all intervals (`cumu prob' in Figure~\ref{fig:parameter_generation}).
Then, we use random values between 0 and 1, which can be mapped to the cumulative probability array utilizing binary search, to select appropriate parameter values for filling the predicate.
In Figure~\ref{fig:parameter_generation}, there are two parameter generation examples.
The complexity of parameter generation is O(log$(H$+$I)$), which is dominated by the binary search.

\iffalse
The data items in HFI of S-Dist are from the workload trace running on real database.
However, the data in synthetic database generated in Section~\ref{sec:databasegeneration} is generally completely different from the data in real database.
For example, the real data $57$ in HFI in Fig.~\ref{fig:basic_distribution} may not exist in our synthetic database.
So before instantiating the symbolic parameters using S-Dist, we need first do data transformation for HFI.
Suppose the column generator for our example is `index = ranInt[1, 400], value = index * 5', where 400 is the column cardinality.
First, we regenerate the data items in HFI using our column generator, as shown in Fig.~\ref{fig:parameter_generation}, where, for example, data item 57 is replaced by 195.
Then, we calculate a cumulative probability array based on the frequencies of high frequency items and all intervals (`cumu prob' in Fig.~\ref{fig:parameter_generation}).
Finally, we use a random generator to generate values between 0 and 1, which can be mapped to the cumulative probability array and used to select appropriate parameter values for filling the predicate.
In Fig.~\ref{fig:parameter_generation}, there are two parameter generation examples.
In this way, we can easily ensure that the generated parameters conform to the desired frequency distribution.
The complexity of parameter generation is O(log$(H$+$I)$), which is dominated by the binary search over the cumulative probability array.
\fi

Additionally, in order to control the cardinality of parameters generated on each interval, we redefine a random index generator for parameter generation, that is `index = $\lfloor$ ranInt[0, cdn$_i$) / cdn$_i$ * cdn$_{avg}$  $\rfloor$ + minIdx$_i$', where cdn$_i$ is the expected cardinality of the generated parameters on the target interval, cdn$_{avg}$ is the average cardinality for each interval, and minIdx$_i$ is the minimum index of the target interval.
The value transformer remains unchanged.
For example in Figure~\ref{fig:parameter_generation}, the random index generator for interval 2 is `index = $\lfloor$ ranInt[0, 50) / 50 * 80  $\rfloor$ + 161', with cdn$_2$ = 50, cdn$_{avg}$ = $400 / 5$ = 80, minIdx$_2$ = $80 * 2 + 1$ = 161.

% Noted that cdn$_i$ may be larger than cdn$_{avg}$, which will cause a certain deviation in the data access distribution of synthetic workloads.
% But for location parameters, especially parameters on primary key columns, this situation is not common because ...

Although the parameter in above example is of integer type, our approach is generic.
For all numeric parameters corresponding to non-key columns, the representation of S-Dist and the parameter generation are exactly the same.
For parameters on key columns, there are small differences.
Since the primary key in our synthetic database is generated sequentially, the domains of key columns in synthetic database may be different from that in real database.
Therefore, when collecting statistics for S-Dist, we use the domains of key columns in the real database to divide the intervals and construct HS.
But during synthetic workload generation, we use the domains of key columns in the synthetic database to support parameter generation.
% , that is, the 'real' domain on each interval is replaced by the 'synthetic' domain.
% Note that for parameter generation corresponding to key columns, index is the output, namely 'value = index'.
For string typed parameters, the biggest difference is how we divide the intervals.
When constructing HS for a string typed parameter, the interval which it belongs to is calculated by $h\%I$, where $h$ is the hash code of the parameter.
The parameter generation is similar to numeric types.
\subsection{Dynamic Data Access Distribution} \label{sec:d-dist}

S-Dist can well depict the skewness of data access distribution throughout the workload cycle.
However, if the data access distribution is dynamically changing, S-Dist is inaccurate or even completely wrong.
Let's take a simple example.
Suppose there is a table with $10^3$ records and a workload with a period of $10^3$ seconds.
In the $i^{th}$ second, all workload requests only access the $i^{th}$ record in the table.
% database requests of the workload
Assume that the database throughput is stable during this period.
% the throughput of the database
At this time, if S-Dist is used to express the entire data access process, we actually find that there is no hot data and the data access distribution is very uniform.
Obviously, this is quite different from the fact.
% For the synthetic workload generated according to this S-Dist, the transaction conflict intensity on database must be much lower than that of the real workload.
So in this section, we propose D-Dist to characterize both dynamics and skewness of data access distribution.

In order to catch such kind of dynamics, we divide the workload trace of a parameter into multiple equal-length time windows according to the log timestamp.
For the parameter trace in any time window, we generate its individual S-Dist, and the D-Dist for entire parameter traces is defined as a list of S-Dists.
Assuming that the entire workload cycle is one day and the time window size is one second, D-Dist of a parameter is constructed by 24*3600=86400 S-Dists.
During the generation of synthetic workloads, we instantiate symbolic parameters using the S-Dist corresponding to the generation time.
% Since the S-Dist is counted for each time window, it should not happen that the hash set consumes excessive memory when collecting the cardinality statistics for each interval.
% Therefore, there is no need to use the Bloom filter or data sampling to obtain approximate cardinality statistics.
In addition, for numeric parameters, the used data range in a time window may be much smaller than column domain.
In order to improve the accuracy of HS, the intervals can be divided according to the data range of current window when collecting the statistics.
Certainly, it is also necessary to use the corresponding index range for each interval when generating the parameters.
% Overall, on the basis of S-Dist, we can easily build the D-Dist for a parameter.

%\vspace{-1mm}
\subsection{Continuous Data Access Distribution} \label{sec:c-dist}
% \vspace{-0.5mm}

% It is well known that the cache hit ratio has an important impact on the database performance.
% Therefore, the data access distribution needs to be able to adequately describe the access characteristics that affect the cache hit ratio.
In some applications, the hot degree of data is closely related to time, and the specific manifestation is that data may be accessed continually for a period of time.
We call this the continuity of data access distribution.
% For example, for online food delivery applications (e.g. ELEME and GrubHub), goods like steamed buns and soybean milk are generally visited frequently in the morning; the gaifan and light meals may be more popular at noon; in the afternoon, coffee and desserts may be people's favorites.
% The same is true for many news applications.
% Data access usually has the property of continuity.
For example, for online food delivery applications, steamed buns are generally ordered frequently in the morning, and coffees may be the favorite one during the afternoon.
Previously, D-Dist is defined to catch the skewness of data access in time windows, while ignoring the continuity of data access between successive time windows.
% In the previously proposed D-Dist, we only considered the skewness of data access in the time window, while ignoring the continuity of data access between time windows.
When using it to generate synthetic workloads, the data accessed between successive time windows may be completely different, which results in a lower cache hit ratio.
% the cache hit ratio in the database may be much lower than that of the real workload, especially when a time window switches to the next time window.
% In this section, we present the Continuous Data Access Distribution (abbreviated C-Dist) that can effectively characterize the continuity, dynamics, and skewness of data access.
So we present C-Dist to characterize the continuity, dynamics, and skewness of data access distribution.

\begin{figure}
	\centering
	\includegraphics[width=0.85\columnwidth]{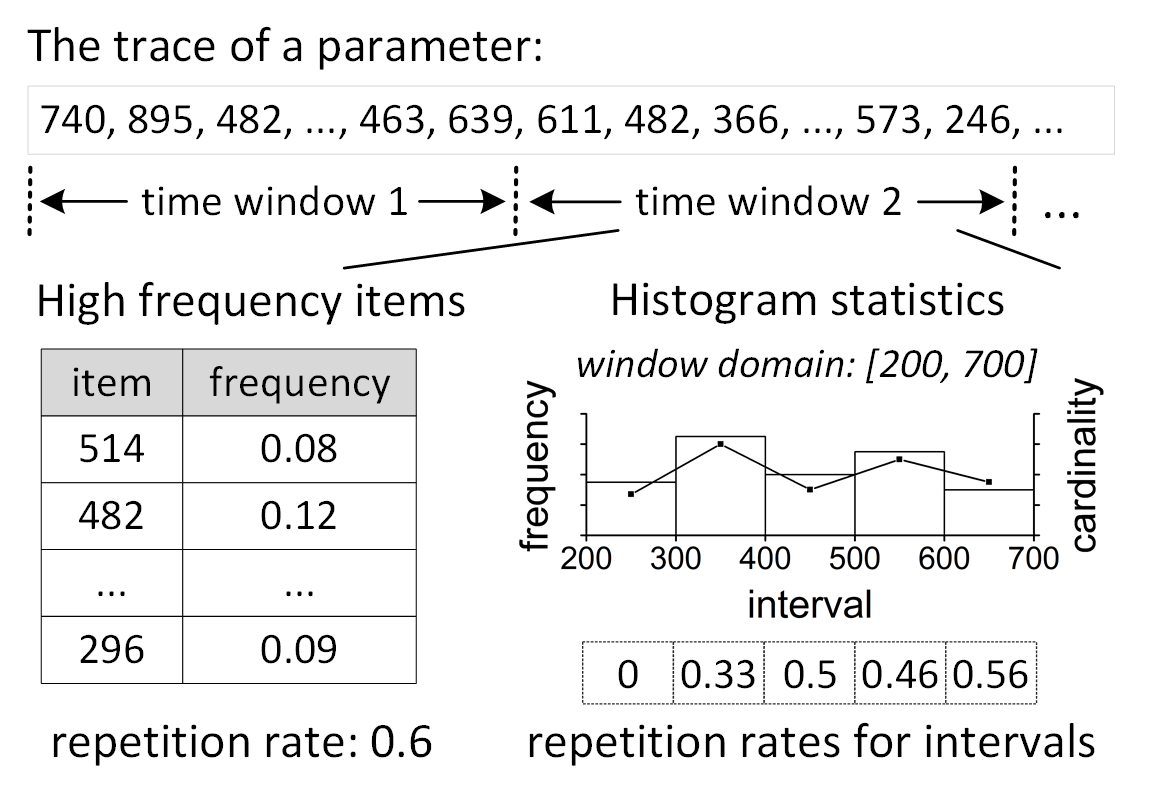}
	%\vspace{-2mm}
	\caption{An example for illustrating C-Dist}
	%\vspace{-3mm}
	\label{fig:cdist}
\end{figure}

We introduce the concept of repetition rate for C-Dist to describe the continuity of data access based on D-Dist.
When collecting the statistics, we count the repetition rate between the high frequency items in current time window and those in previous time window, as well as the repetition rates of parameters for all intervals.
Figure~\ref{fig:cdist} continues the example in Figure~\ref{fig:basic_distribution} by adding repetition rates for both HFI and HS.
In this example, the repetition rate for HFI is 0.6, that is, there are three high frequency items are kept from the previous time window.
The repetition rates for five intervals in HS are 0, 0.33, 0.5, 0.46 and 0.56.
Assuming that cdn$_1$ is 15, then there are $15*0.33\approx5$ parameter values in interval 1 that have appeared in the previous time window.
% For the first time window, all repetition rates are 0.
% Currently, we only consider the continuity of data access between the successive time windows.
% The continuity over multiple time windows is left as a future work.

% \vspace{-1mm}
\begin{algorithm}[h]
	\renewcommand{\algorithmicrequire}{\textbf{Input:}}
	\renewcommand{\algorithmicensure}{\textbf{Output:}}
	\caption{ Candidate parameter generation }
	{\small 
		\begin{algorithmic}[1]
			\REQUIRE High frequency items $\Phi^{t-1}$ and candidate parameters for intervals $\Omega^{t-1}$ of previous time window, repetition rate $\kappa$ for HFI, repetition rates $\Psi$ for HS, cardinalities $\mathcal{C}$ in HS for intervals, number of high frequency items $H$, number of intervals $I$
			\ENSURE High frequency items $\Phi^t$ and candidate parameters for intervals $\Omega^t$ of current time window
			
			\STATE Randomly select $H*\kappa$ items from $\Phi^{t-1}$ and put them in $\Phi^t$.
			\STATE Randomly generate $H*(1-\kappa)$ items with the corresponding column generator and put them in $\Phi^t$. % All items in $\Phi$ are guaranteed to be different.
			
			\STATE Initialize an array $\Gamma \leftarrow 0...0$. The length of $\Gamma$ is $I$.
			\STATE Initialize a list $P \leftarrow$ all parameters in $\Omega^{t-1}$. % And shuffle $P$.
			
			\FORALL{ parameter $p$ in $P$ }
			\STATE Identify the interval number $i$ of parameter $p$.
			\IF{ $i$ is valid $\&\&$ $\Gamma[i] < \mathcal{C}[i] * \Psi[i]$ }
			\STATE Put the parameter $p$ in $\Omega^t[i]$, $\Gamma[i]++$
			\ENDIF\ // {\footnotesize $i$ is invalid when $p$ is not in the current window domain.}
			\ENDFOR
			
			\FORALL{ interval $i = 0$ to $(I-1)$ }
			% \STATE Randomly generate $(\mathcal{C}[i]-\Gamma[i])$ parameters belonging to interval $i$ using the corresponding column generator and put them in $\Omega[i]$. The $(\mathcal{C}[i]-\Gamma[i])$ parameters added are guaranteed to be different from those in $P$.
			\STATE Randomly generate $(\mathcal{C}[i]-\Gamma[i])$ parameters belonging to the interval $i$ using the column generator and put them in $\Omega^t[i]$.
			\ENDFOR
			
			\RETURN $\Phi^t$ and $\Omega^t$
		\end{algorithmic} 
	}\label{alg:candi_para_gene}
\end{algorithm}
% \vspace{-5mm}

In order to ensure the repetition rates in C-Dist, we need to pregenerate the candidate parameters for each time window.
In Algorithm~\ref{alg:candi_para_gene}, we list the detailed generation process of candidate parameters.
In lines 1-2, we generate all the high frequency items that satisfy the expected repetition rate. 
In lines 3-10, we traverse all the parameters in previous time window and select the repeated parameters for each interval of current time window until the repetition rate on the interval is met.
% In the $6^{th}$ line, we deduce the index of the parameter according to its value, so as to identify the interval number which it belongs to in the current time window.
% If the index is not in the index domain of the current time window, the parameter is ignored.
% And for string typed parameters, e.g., `{\em $296$\#$dgtckuy$}', the front part of the value split by `\#' character is the index we need.
Finally, in lines 11-13, we generate random parameters added to each interval to reach the cardinality requirement.
Now for parameter generation against a certain interval, we only need to randomly select a candidate parameter as the output.
% Based on these candidate parameters, our parameter generation mechanism is still consistent with Fig.~\ref{fig:parameter_generation}.
% And now for a certain interval, we only need to randomly select a candidate parameter as the output.
% Last but not least, 
In addition, if the candidate parameters are generated online during the synthetic workload generation, the workload generator may become the performance bottleneck, thereby affecting the correctness of evaluation results.
Therefore, we can generate the candidate parameters for all time windows offline and store them on disk, and then read them as needed when generating the synthetic workloads.

\iffalse
{\color{blue}{
		\textbf{Discussion.}
		The setting of the time window size depends on how often the target workload changes.
		If the workload changes frequently, the window size should be set to a small value, otherwise it can be set to a larger value.
		% For instance, when the time window size is set to 1 second, we can capture the seconds-level workload change.
		We recommend setting the time window size to 1 second so that even second-level workload changes can be captured, which is enough for most applications.
		Histogram is a commonly used statistical method for density estimation, which are widely used to represent data distribution statistics in industry databases, such as Oracle, MySQL and PostgreSQL. % ~\cite{b30Springer2013}
		And we can apply academic and industrial mature methods~\cite{b30ESAIMPS2006, b31OracleHistograms} to choose the number of intervals.
}}\fi
\vspace{0.5mm}
\section{Workload Generation} \label{sec:workloadgeneration}
\vspace{0.5mm}

% Given the transaction logic of each transaction and the data access distribution of each parameter, the workload generator is responsible for generating synthetic workloads satisfying the desired workload characteristics and configurations.

Given the transaction logic of each transaction template and the data access distribution of each parameter, the {\em Workload Generator} in Figure~\ref{fig:lauca-arch} is responsible for generating synthetic workloads satisfying the desired configurations.
% Meanwhile, efficiently generating high concurrency/throughput synthetic workloads in a distributed environment is also an essential requirement for our workload generator.
% Now we will 
We present the details of workload generation from three levels: thread model, transaction execution and parameter instantiation.

% \vspace{+1mm} \noindent
\textbf{Thread Model.}
Users can configurate multiple test nodes for deploying the workload generator and the number of test threads on each node.
% , to simulate the concurrency.
% Based on this, we start a JVM process and the specified number of test threads (i.e. test clients) on each test node, and establish a separate database connection for each test thread.
For each test thread, we establish a separate database connection.
The concurrency of synthetic workload is the number of all test threads.
Lauca supports two different execution models for test threads to invoke transactions: {\em no-await-in-loop} and {\em fixed-throughput}.
% Allowing users to choose which execution model to use in their testing enables them to build extended synthetic workloads.
With the no-await-in-loop setting, all test threads repeatedly issue transactions without any think time between requests.
In fixed-throughput setting, the user can specify a fixed request throughput or a throughput scale factor.
If the throughput scale factor is specified, we multiply the scale factor and the throughput of each time window obtained from workload traces to get the target throughput.
The test threads achieve the desired throughput by controlling the think time between transaction requests.
% 这句可以注释~
% When the required throughput exceeds the maximum throughput that the current test threads can achieve, the execution model degenerates to no-await-in-loop.
Different execution models enable us to build extended synthetic workloads.

% \vspace{+1mm}
% \noindent
\textbf{Transaction Execution.}
The test thread invokes different kinds of transactions according to their proportions extracted from workload traces.
And the transaction proportions are adjusted periodically with the time window.
% Users can specify whether the operations in a transaction are executed in a precompiled manner.
During the execution of a transaction, the structure information of its transaction logic will be used to determine whether the operations in the branch structure need to be executed, and the number of executions of operations in the loop structure.
For the execution of a specific SQL operation, we first instantiate all the symbolic parameters one by one, and then send the operation with concrete parameter values to the test database.
After the operation is executed, the result set and the parameters are maintained in an intermediate state for the generation of other parameters in subsequent operations within the same transaction instance.

% \vspace{+1mm}
% \noindent
\textbf{Parameter Instantiation.}
When instantiating the parameters, we first guarantee the consistency of transaction logic, and then ensure the data access distribution of synthetic workloads.
% The parameter dependencies and the data access distribution of a parameter are used to support the corresponding parameter instantiation.
% And parameter dependencies are used preferentially to instantiate the parameter for promising the consistent business logic.
% As the transaction logic definition in Section~\ref{sec:txlogicdefinition}, there are three types of parameter dependencies.
% \ding{202} 
For a parameter, {\em Case 1: if there is only {\small PD}1,} the value of this parameter can be calculated directly based on the increment $\Delta$ and the value of dependent parameter.
% And there is certainly no other parameter dependency here.
% In fact, the data access distribution of this parameter does not need to be counted.
% \ding{203} 
{\em Case 2: if there is only {\small PD}2,} we first attempt to instantiate the parameter by randomly selecting a dependency according to their probabilities.
When no dependencies are selected, then we use the data access distribution to instantiate this parameter.
% For example, suppose that the parameter has two dependencies, and the probabilities are 0.5 and 0.3, respectively.
% At this point, the probabilities that two dependencies are used to instantiate the parameter are, of course, 0.5 and 0.3, respectively, and there is a probability of 0.2 using the data access distribution to instantiate the parameter.
% \ding{204} 
{\em Case 3: if there are both {\small PD}2 and {\small PD}3,} the corresponding operation must be in the loop structure.
% 3) Moreover, there may be both dependencies \$2 and \$3 for the parameters of the operation in the loop structure.
For the first loop execution, we still instantiate the parameter based on {\small PD}2 and data access distribution as Case 2.
For non-first loop executions, we first attempt to instantiate the parameter using {\small PD}3 based on the probability.
If no dependency in {\small PD}3 is selected, {\small PD}2 and data access distribution are then used to instantiate the parameter.
% \ding{205} Finally, if the parameter does not have any dependencies (e.g. the first parameter in the transaction template), we can directly use the data access distribution to instantiate the parameter.

% 序号可以换成好看点的字符表示

Overall, transaction execution and parameter instantiation are independent of each other for all test threads, so our workload generator can be deployed on multiple nodes to efficiently generate the high concurrency/throughput synthetic workloads while satisfying the desired workload characteristics and configurations.

\section{Experiments} \label{sec:experiments}
%\vspace{1mm}

\begin{figure*}
	\centering
	\subfigure[Throughput]{
		\label{fig:tpcc-pg-throughput}
		\includegraphics[width=0.23\linewidth]{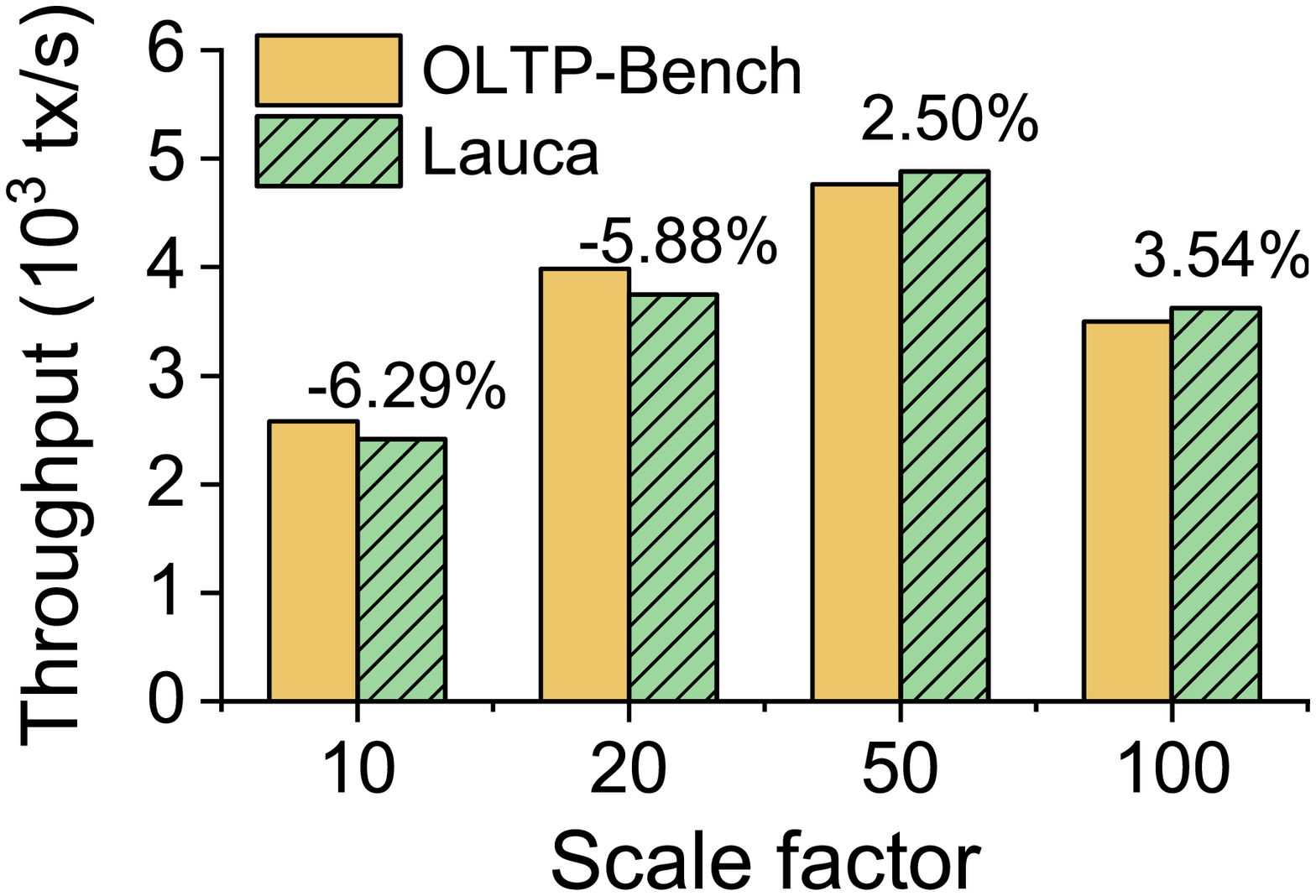}
	}
	\subfigure[Latency]{
		\label{fig:tpcc-pg-latency}
		\includegraphics[width=0.23\linewidth]{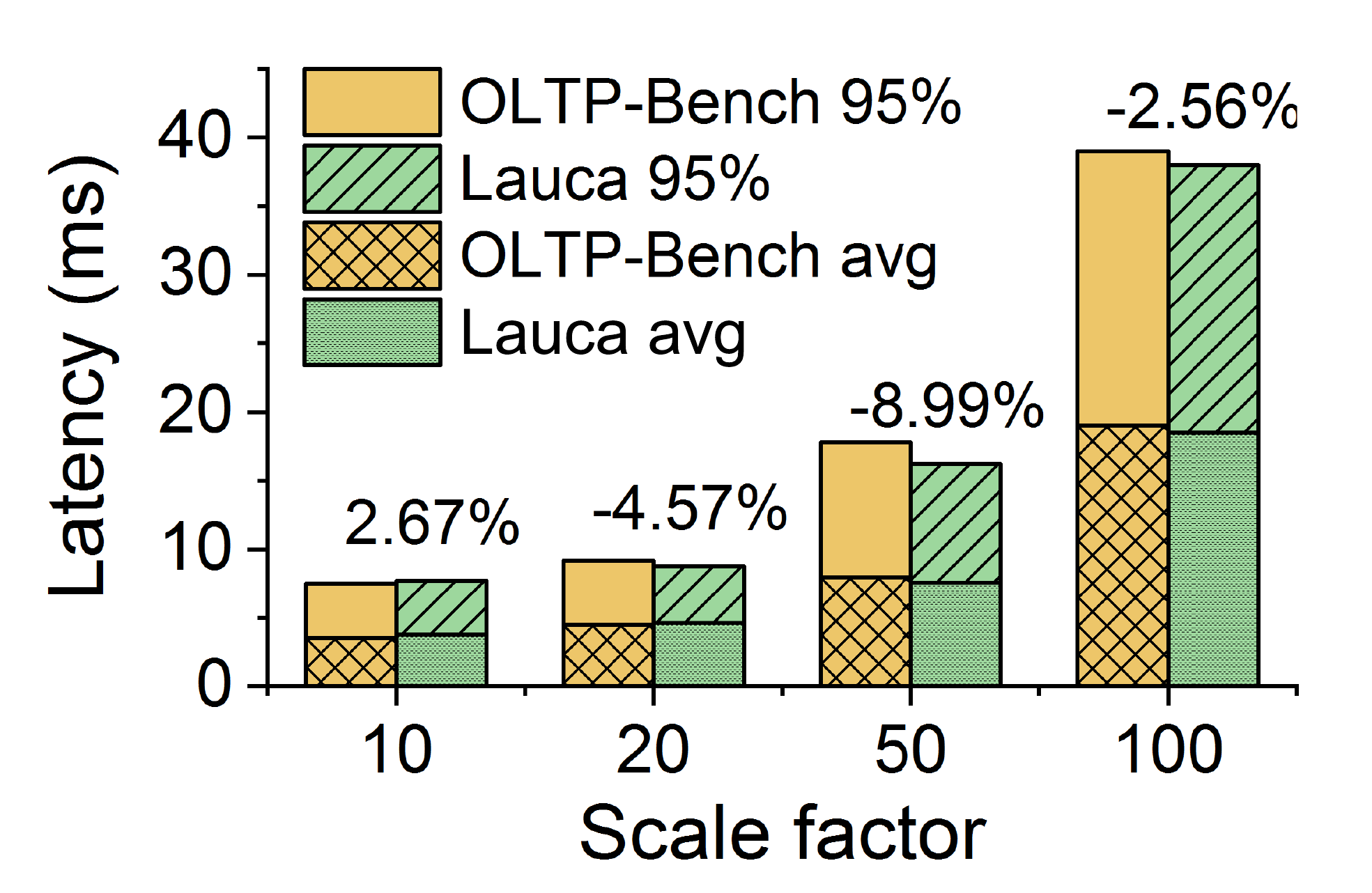}
	}
	\subfigure[CPU usage]{
		\label{fig:tpcc-pg-cpu}
		\includegraphics[width=0.23\linewidth]{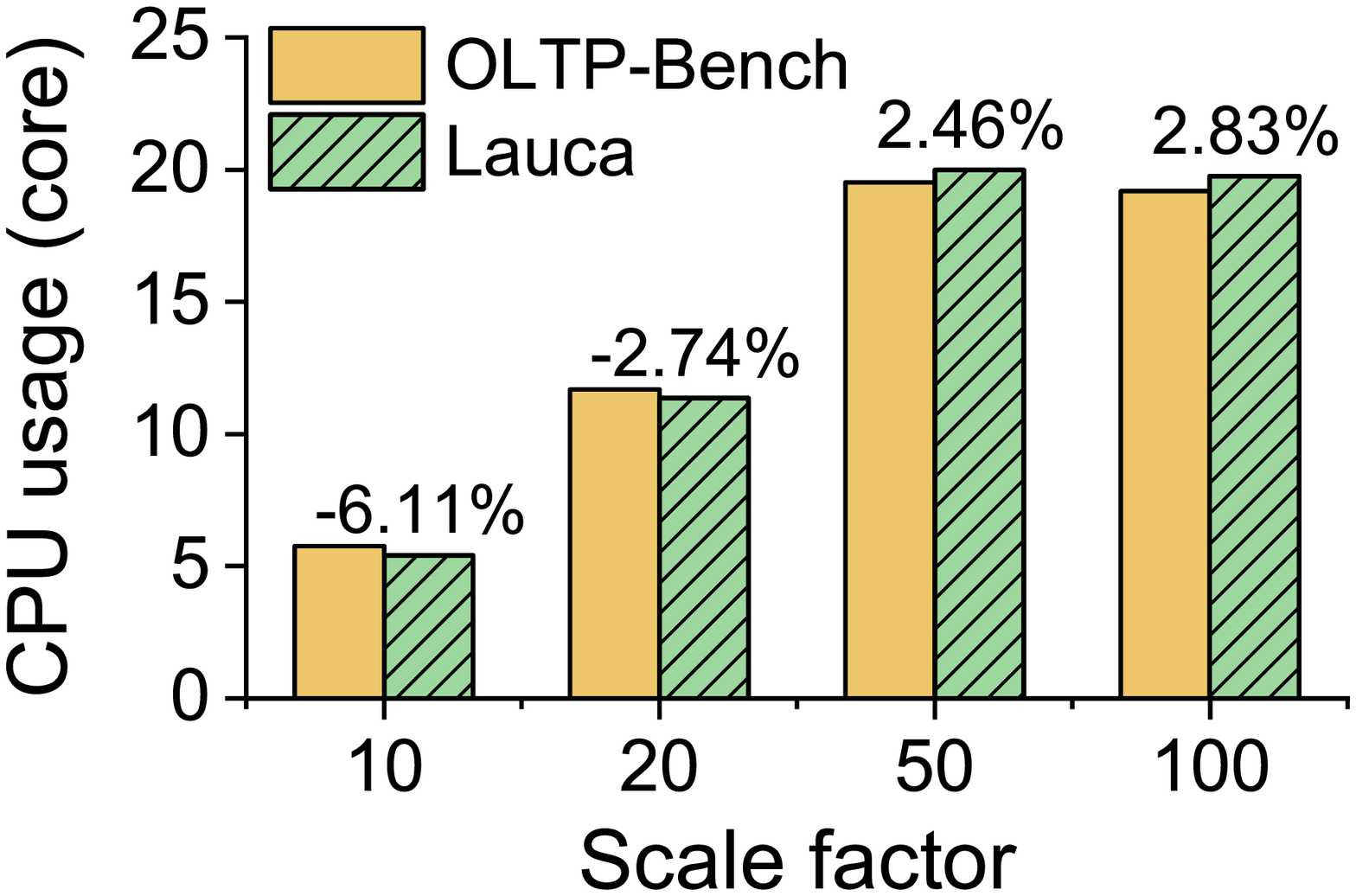}
	}
	\subfigure[Disk usage]{
		\label{fig:tpcc-pg-disk}
		\includegraphics[width=0.23\linewidth]{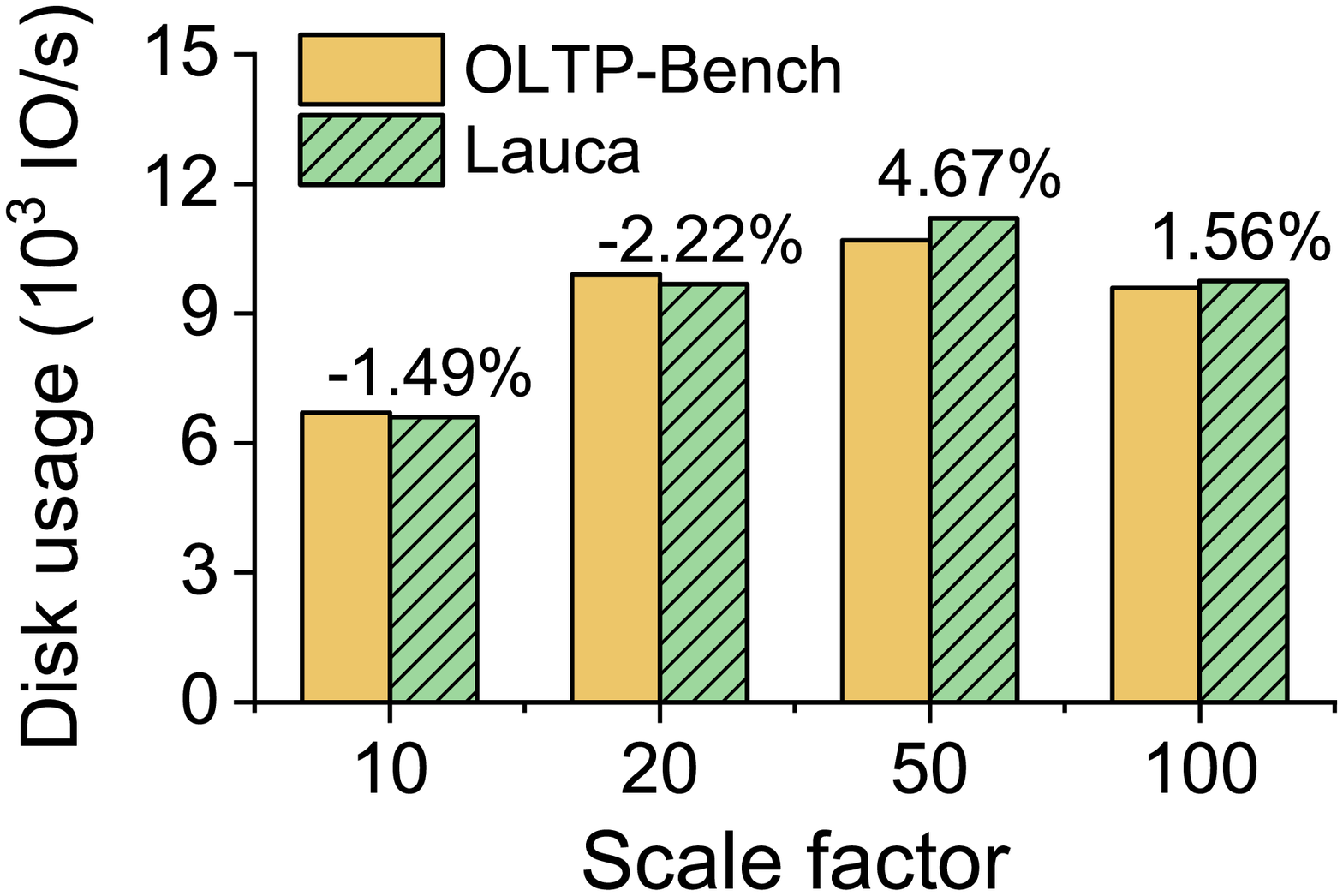}
	}
	%\vspace{-4.5mm}
	\caption{Deviations in performance metrics for TPC-C workloads on PostgreSQL database}
	%\vspace{-3mm}
	\label{fig:tpcc-pg}
\end{figure*}

\begin{figure*}
	\centering
	\subfigure[Throughput]{
		\label{fig:smallbank-mysql-throughput}
		\includegraphics[width=0.23\linewidth]{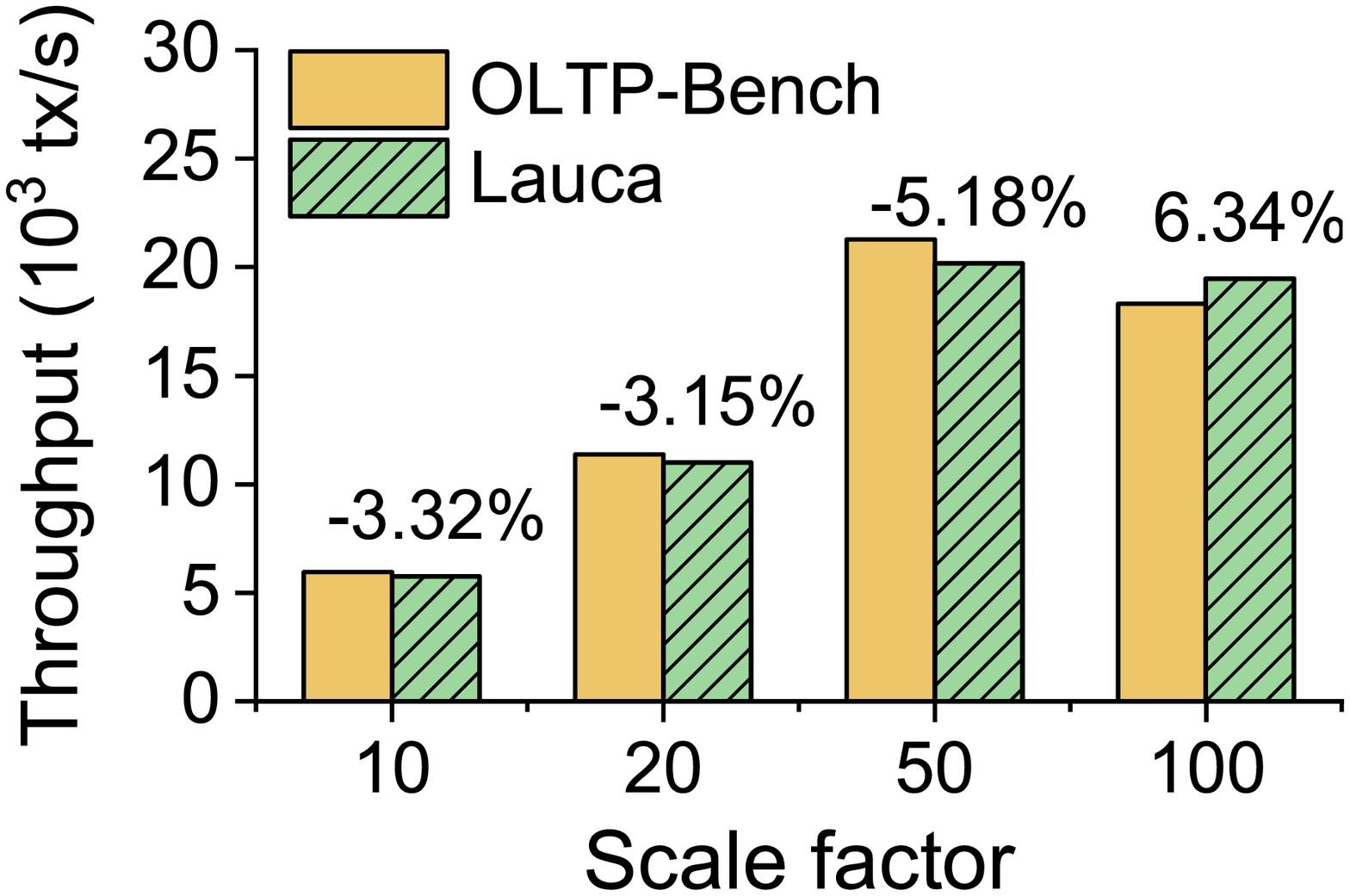}
	}
	\subfigure[Latency]{
		\label{fig:smallbank-mysql-latency}
		\includegraphics[width=0.23\linewidth]{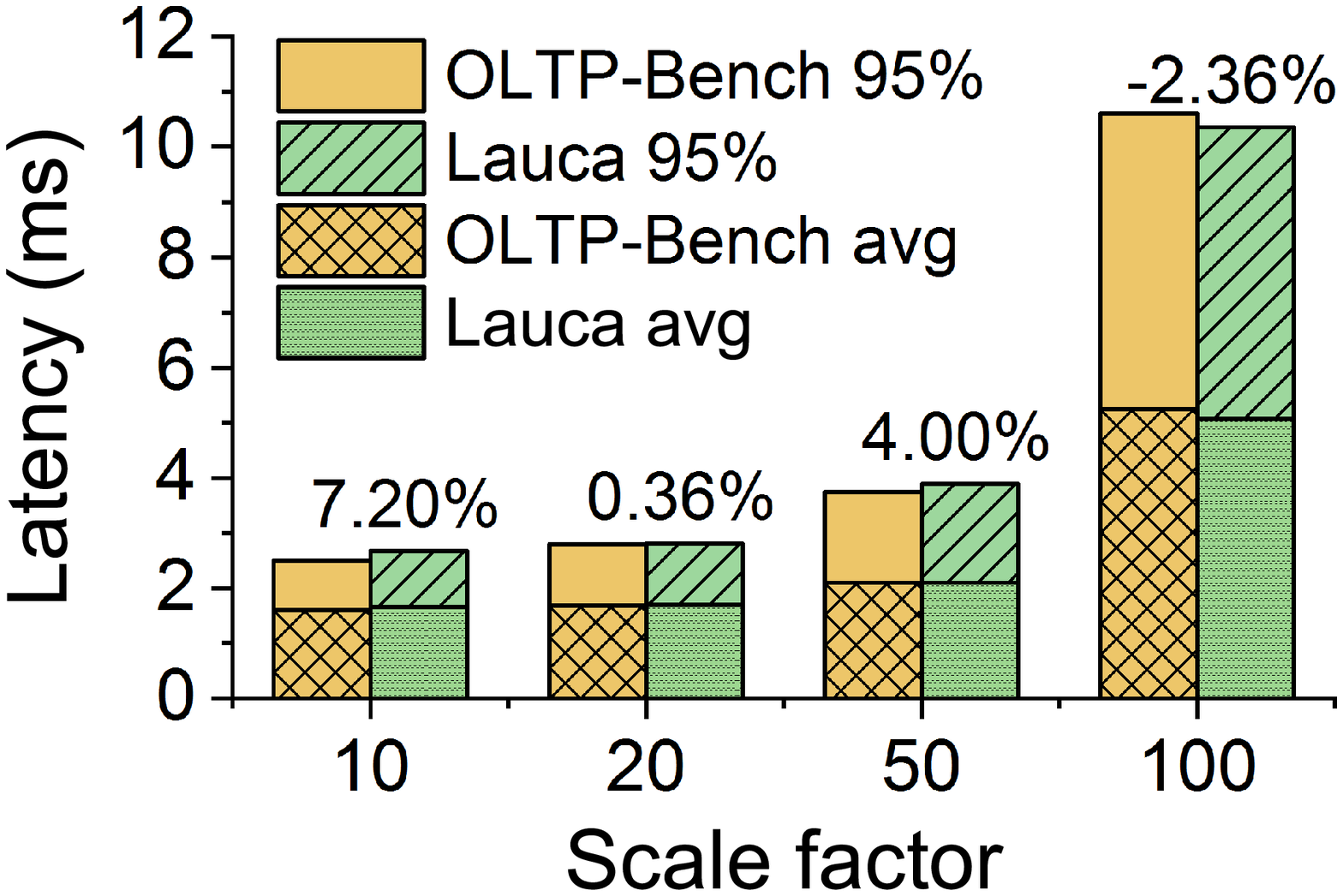}
	}
	\subfigure[CPU usage]{
		\label{fig:smallbank-mysql-cpu}
		\includegraphics[width=0.23\linewidth]{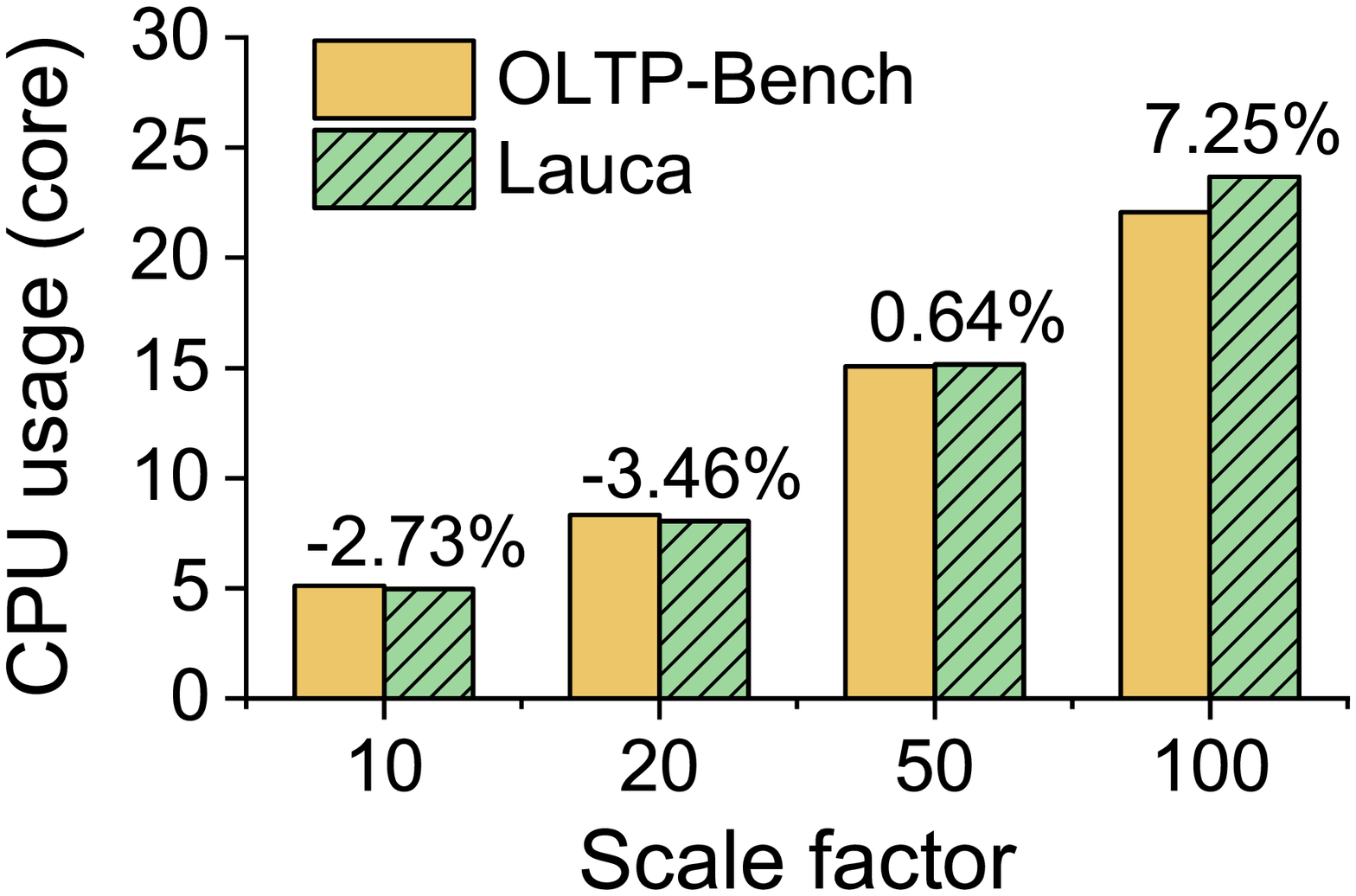}
	}
	\subfigure[Disk usage]{
		\label{fig:smallbank-mysql-disk}
		\includegraphics[width=0.23\linewidth]{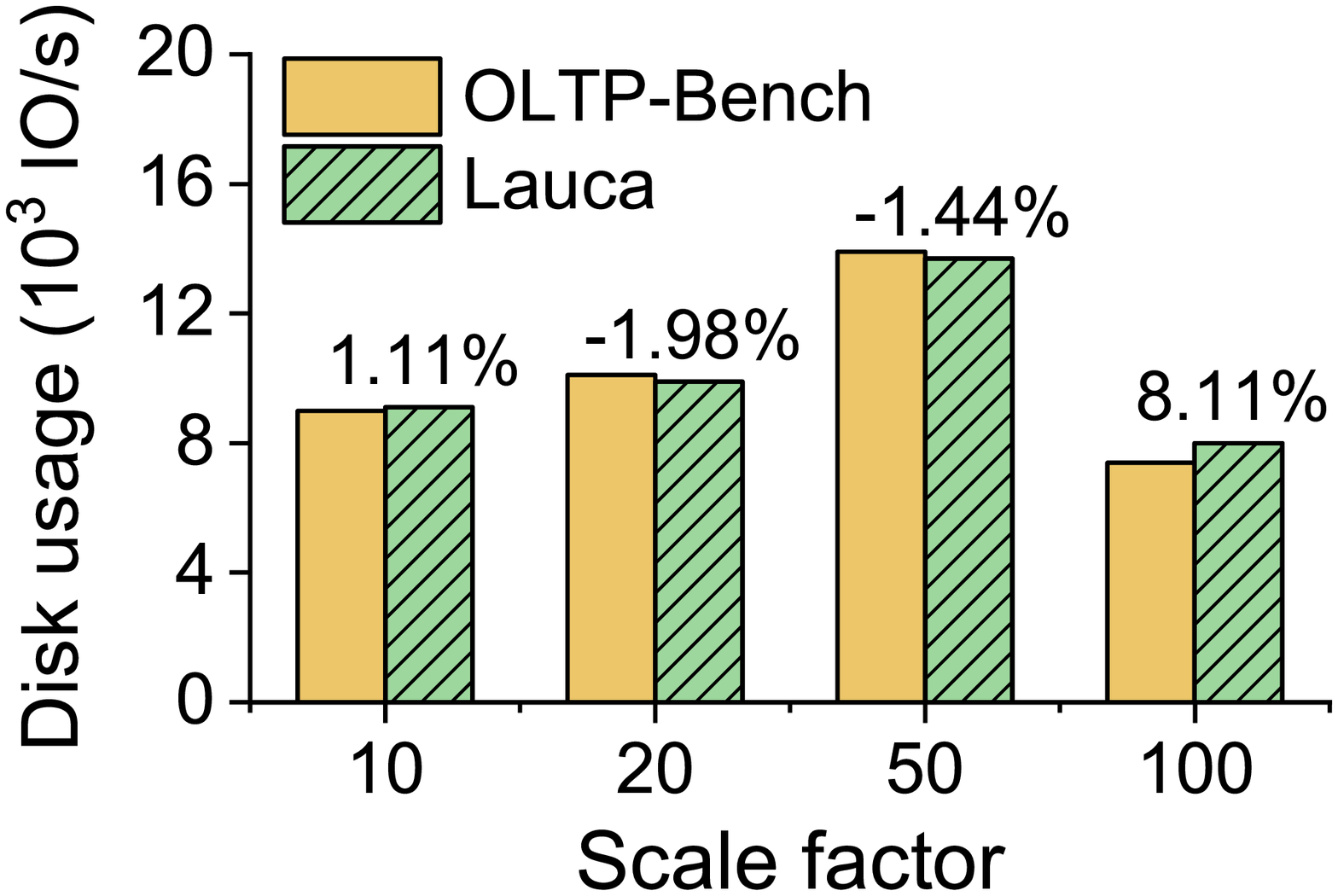}
	}
	%\vspace{-4.5mm}
	\caption{Deviations in performance metrics for SmallBank workloads on MySQL database}
	%\vspace{-1mm}
	\label{fig:smallbank-mysql}
\end{figure*}

% 实验安排：
% 最终确定的负载有：tpcc，smallbank，ycsb（扩展）
% 最终确定的数据库有：MySQL，PostgreSQL

{\bfseries Environment.}
Our experiments are conducted on four servers, and each server is equipped with 2 Intel Xeon Silver 4110 @ 2.1 GHz CPUs, 120 GB memory, 4 TB HDD disk configured in RAID-5 and 4 GB RAID cache.
The servers are connected using 10 Gigabit Ethernet.
% 'sudo megacli -AdpAllInfo -aALL' for getting RAID cache
The test platforms selected in our experiments are the most popular and advanced open source DBMSs: MySQL (v5.7.24) and PostgreSQL (v10.4), which are also widely used in industry.

{\bfseries Workloads.}
Two standard benchmarks are used throughout the experiments: TPC-C~\cite{tpcc} and SmallBank~\cite{alomari2008cost}.
TPC-C is one of the most widely used industrial-level OLTP benchmark. % online transaction processing
It involves nine tables and five types of transactions that simulate the activities found in complex OLTP applications.
SmallBank abstracts the operations in banking applications, and it includes three tables and six types of transactions.
% where the schema consists of three tables and the workload includes six types of transactions.
All transactions of SmallBank perform simple read and update operations on a small number of tuples.
We use OLTP-Bench~\cite{difallah2013oltp}, an extensible DBMS benchmarking testbed, to generate workloads of TPC-C and SmallBank, which act as real workloads for comparison.
Workload traces are logged by OLTP-Bench and serve as input to Lauca for generating the synthetic workloads.
By the way, the original TPC-C implementation of OLTP-Bench artificially binds the test threads to the warehouses, which greatly reduces the conflict of generated workloads on the database.
This additional binging is removed in our experiments to make the workload more practical.
Yahoo! Cloud Serving Benchmark (YCSB)~\cite{cooper2010benchmarking} is a collection of workloads that represent the large-scale web applications.
% and are mainly single-row transactions.
We construct micro benchmark workloads based on YCSB, for simulating workloads with fine-grained control on skewness, dynamics and continuity.

{\bfseries Settings and Setup.}
In the transaction logic extraction, the number $K$ of transaction instances and the group number $N$ of transaction instances are all set to $10^4$.
For the data access distribution extraction, the number $H$ of items in HFI and the number $I$ of intervals in HS are all set to 50, and the time window size is set to 1 second.
Both MySQL and PostgreSQL are deployed on a single server, respectively.
By default, Lauca/OLTP-Bench/YCSB is deployed on a single server that does not deploy any database system.

% MySQL和PostgreSQL数据库分别部署在一个单节点上，默认情况下，Lauca被部署在一个没有数据库服务的单节点上。

\iffalse
\begin{figure*}
	\centering
	\subfigure[Throughput]{
		\label{fig:tpcc-mysql-throughput}
		\includegraphics[width=0.231\linewidth]{figures/tpcc-mysql-throughput.eps}
	}
	\subfigure[Latency]{
		\label{fig:tpcc-mysql-latency}
		\includegraphics[width=0.231\linewidth]{figures/tpcc-mysql-latency.eps}
	}
	\subfigure[CPU usage]{
		\label{fig:tpcc-mysql-cpu}
		\includegraphics[width=0.231\linewidth]{figures/tpcc-mysql-cpu.eps}
	}
	\subfigure[Disk usage]{
		\label{fig:tpcc-mysql-disk}
		\includegraphics[width=0.231\linewidth]{figures/tpcc-mysql-disk.eps}
	}
	\caption{Deviations in performance metrics for TPC-C workloads on MySQL database}
	\label{fig:tpcc-mysql}
\end{figure*}
\fi

\iffalse
\begin{figure*}
	\centering
	\subfigure[Throughput]{
		\label{fig:smallbank-pg-throughput}
		\includegraphics[width=0.231\linewidth]{figures/smallbank-pg-throughput.eps}
	}
	\subfigure[Latency]{
		\label{fig:smallbank-pg-latency}
		\includegraphics[width=0.231\linewidth]{figures/smallbank-pg-latency.eps}
	}
	\subfigure[CPU usage]{
		\label{fig:smallbank-pg-cpu}
		\includegraphics[width=0.231\linewidth]{figures/smallbank-pg-cpu.eps}
	}
	\subfigure[Disk usage]{
		\label{fig:smallbank-pg-disk}
		\includegraphics[width=0.231\linewidth]{figures/smallbank-pg-disk.eps}
	}
	\caption{Deviations in performance metrics for SmallBank workloads on PostgreSQL database}
	\label{fig:smallbank-pg}
\end{figure*}
\fi

%\vspace{-1mm}
\subsection{Fidelity of Synthetic Workloads} \label{sec:workloadfidelity}

The similarity between synthetic workloads and real application workloads is called workload {\em fidelity}, which is the most essential target when designing Lauca.
% Fidelity of synthetic workloads is the most essential target when designing Lauca.
It is measured by performance deviations obtained by running synthetic workloads and real application workloads on the same database system.
% We take the most critical performance metrics for database evaluations, which are throughput, average latency, 95\% latency, CPU and Disk IO usage.
% Small deviations indicate high similarity between the synthetic workload and the real one, which is much more useful for database performance testing.
% The test platforms selected in our experiments are the most popular and advanced open source DBMSs: MySQL (v5.7.24) and PostgreSQL (v10.4), which are also widely used in industry.

\iffalse
The fidelity of synthetic workloads is the most essential guarantee for {\em Lauca}'s effectiveness.
Fidelity can be measured by exploiting the deviations of performance metrics obtained from the running of synthetic workload and real application workload on the same database system.
And the small deviation indicates that the synthetic workload is similar to the real application workload so as to provide insight in database performance testing.
Due to the space constraint, only some of the most critical performance metrics are shown in the experimental results, and they are throughput, average latency, 95\% latency, CPU and Disk IO usage.
The test platforms selected in our experiments are the most popular and advanced open source DBMSs: MySQL (v5.7.24) and PostgreSQL (v10.4), which are widely used in industry.
\fi

Figure~\ref{fig:tpcc-pg}-\ref{fig:smallbank-mysql} show the deviations in performance metrics between the synthetic workloads generated by Lauca and the real workloads generated by OLTP-Bench, under different scale factors.
The concurrency of database requests is the same as the scale factor.
There are two groups of experiments, respectively, for the executions of TPC-C workloads on PostgreSQL database and SmallBank workloads on MySQL database.
% There are four groups of experiments, respectively, for the executions of the TPC-C or SmallBank workloads on the MySQL or PostgreSQL database systems.
% Among them, Figure~\ref{fig:tpcc-mysql} is for the TPC-C workloads and MySQL database.
In Figure~\ref{fig:tpcc-pg-throughput} and Figure~\ref{fig:smallbank-mysql-throughput}, we present the transaction execution throughputs of real workloads and synthetic workloads. % with different benchmarks on different databases.
% The concurrency of database requests is the same as the scale factor, and that is true for all the experiments in Figures~\ref{fig:tpcc-pg}-\ref{fig:smallbank-mysql}.
% Database system (here is MySQL) and test clients (i.e. OLTP-Bench or {\em Lauca}) are deployed at different nodes.
From the results we can see that the throughputs of the two workloads are very similar, and the biggest deviation is as low as 6.29\% in Figure~\ref{fig:tpcc-pg-throughput} and 6.34\% in Figure~\ref{fig:smallbank-mysql-throughput} respectively.
For average latency and 95\% latency metrics, in Figure~\ref{fig:tpcc-pg-latency} and Figure~\ref{fig:smallbank-mysql-latency}, we can see that the synthetic workload is very close to the real workload on both two metrics, with the maximum deviation between the two workloads is only 8.99\% and 7.20\% respectively.
Figure~\ref{fig:tpcc-pg-cpu}-\ref{fig:tpcc-pg-disk} and Figure~\ref{fig:smallbank-mysql-cpu}-\ref{fig:smallbank-mysql-disk} report the CPU and Disk usages of the two workloads.
The results show that the resource consumptions for executing synthetic workloads and real workloads on both PostgreSQL and MySQL databases are consistent, which further verify the high fidelity of synthetic workloads generated by Lauca.

\begin{table}[H]
	\small
	\centering
	%\vspace{-2mm}
	\caption{Maximum deviations in performance metrics} \label{tab:maxdeviations}
	%\vspace{-2mm}
	\begin{tabular}{|c|c|c|} \hline
		& TPC-C on MySQL & SmallBank on PostgreSQL \\ \hline
		Throughput & 7.35\% & 9.44\% \\ \hline
		Latency & -3.11\% & -8.63\% \\ \hline
		CPU usage & -5.75\% & 7.50\% \\ \hline
		Disk usage & -2.86\% & 7.33\% \\ \hline
	\end{tabular}
	%\vspace{-2mm}
\end{table}

We present the maximum deviations in performance metrics for TPC-C workloads on MySQL database and SmallBank workloads on PostgreSQL database in Table~~\ref{tab:maxdeviations}.
The experimental results are the same as in Figure~\ref{fig:tpcc-pg}-\ref{fig:smallbank-mysql}.
Overall, whether for complex workloads of TPC-C or simple workloads of SmallBank on MySQL or PostgreSQL database, the synthetic workloads generated by Lauca are consistent with the real workloads from OLTP-Bench on various performance metrics.
It conforms that the synthetic workload generated by Lauca has extremely high fidelity.

% Since Lauca has almost the same fidelity on both MySQL and PostgreSQL, we use MySQL for the following experiments to show influence of different components to the performance.

\subsection{Exploring Transaction Logic} \label{sec:txlogicexp}

Simulation of transaction logic of application workloads is an important feature of Lauca.
In this section, we demonstrate the impact of transaction logic on transaction semantics, transaction conflict intensity, deadlock possibility and distributed transaction ratio of synthetic workloads.
All experiments in this section are run on MySQL database, with the workloads taken from TPC-C benchmark.
Both the scale factor and request concurrency are 20.
Though transaction logic consists of structure information and parameter dependency information, in which structure information, such as the number of loop executions, has an obvious impact on performance, so here we mainly explore the impact of parameter dependency information on database performance.

%\vspace{-3mm}
\begin{figure}[H]
	\centering
	\includegraphics[width=0.62\columnwidth]{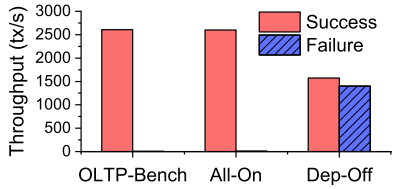}
	%\vspace{-3mm}
	\caption{The impact on transaction semantics}
	%\vspace{-3mm}
	\label{fig:tpcc-txlogic-5tx}
\end{figure}

Using the TPC-C workload including all five types of transactions, we study how the transaction logic affects the transaction semantics of synthetic workloads.
Figure~\ref{fig:tpcc-txlogic-5tx} shows the throughputs of real workloads and synthetic workloads respectively generated by OLTP-Bench and Lauca.
% Both the scale factor and request concurrency are 20.
The transaction throughput is divided into two parts which are {\em success throughput} and {\em failure throughput}.
Success throughput refers to the throughput of successfully executed transactions, while failure throughput refers to the throughput of failed transactions.
In Figure~\ref{fig:tpcc-txlogic-5tx}, when we use all the information in transaction logic (All-On), Lauca presents excellent performance similarity to the real workload; when we turn off the parameter dependency information (Dep-Off), transaction failures increase sharply and the success throughput is much smaller than that of OLTP-Bench.
This is because the insert operations in NewOrder transactions do not satisfy the primary/foreign constraints, leading to a large number of transaction rollbacks.
And because no new order is generated, Delivery transactions cannot be successfully executed too.
Overall, the transaction logic can effectively ensure that the transaction semantics of the synthetic workload are consistent with the real workload.

%\vspace{-2.5mm}
\begin{figure}[H]
	\centering
	\includegraphics[width=0.98\columnwidth]{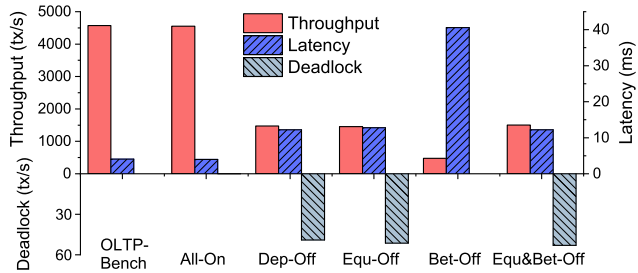}
	%\vspace{-2mm}
	\caption{The impact on transaction conflict intensity, deadlock possibility and scan data volume}
	%\vspace{-3.5mm}
	\label{fig:tpcc-txlogic-3tx}
\end{figure}

To further explore the impact of transaction logic on other aspects, we focus on three types of transactions that can be successfully executed, namely the Payment, OrderStatus and StockLevel transactions. % in TPC-C workloads. % , in Fig.~\ref{fig:tpcc-txlogic-3tx}
In Figure~\ref{fig:tpcc-txlogic-3tx}, we present the transaction throughputs, latencies and deadlock throughputs of workloads generated by OLTP-Bench and Lauca.
There are five groups of experimental results for Lauca, which are obtained by turning off different parts of transaction logic.
Note that {\em latency} here is the average latency of successfully executed transactions. %, not including the failed ones.
From the results, we can see that when we turn off the equal parameter dependency (Equ-Off), the throughput decreases significantly, the latency increases markedly, and a lot of deadlocks occur.
Increased latency indicates that there is more lock waiting time and the transaction conflict is more intensive.
Occurrence of these phenomena is because there are three pairs of read and write operations involving the same record in Payment transactions.
% When the equal parameter dependency is turned off
Under the Equ-Off, read and write operations on the same record are likely to become on different records, which significantly increase the likelihood of transaction conflicts and deadlocks. % greatly, especially when these records are in a small table.
When we turn off the between parameter dependency (Bet-Off), the latency increases drastically, and the throughput is very low.
This is because the scan operation in StockLevel transactions will involve a large amount of data, and normally only about 20 records should be accessed.
% Two parameters that bound the scan operation range, one of which is determined by the equal dependency and the other by the between dependency.
When we turn off both the equal and between dependencies, the scan operation does not read a large amount of data due to the role of C-Dist, so that the performance metrics of Dep-Off, Equ\&Bet-Off and Equ-Off are similar.
Overall, the results confirm that manipulating transaction logic enables synthetic workloads with the same transaction conflict intensity, deadlock possibility and scan data volume as real workloads.

%\vspace{-3mm}
\begin{figure}[H]
	\centering
	\includegraphics[width=0.62\columnwidth]{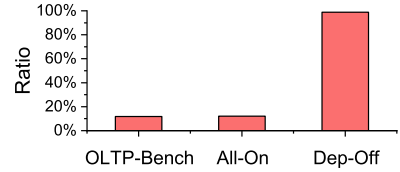}
	%\vspace{-3mm}
	\caption{The impact on distributed transaction ratio}
	%\vspace{-3mm}
	\label{fig:tpcc-txlogic-dtx}
\end{figure}

% The overhead of processing distributed transactions can degrade the system performance significantly~\cite{b12PVLDB2017, b7SIGMOD2006, b13PVLDB2010}.

% The distributed transaction ratio has a significant impact on database performance~\cite{b12PVLDB2017, b7SIGMOD2016, b13PVLDB2010}.
% In order to accurately obtain the distributed transaction ratios of evaluation workloads, we assume that the data is hash-partitioned by Warehouse ID on five nodes, and count the distributed transaction ratios of workloads from the application side.

The distributed transaction ratio has been proved to have a significant impact on database performance in many works~\cite{harding2017evaluation, lin2016towards, curino2010schism}.
We simulate a distributed environment on the single-node MySQL database by assuming that data is hash-partitioned into five virtual nodes according to Warehouse ID.
If the data in a transaction involves multiple virtual nodes, then the transaction is considered a distributed transaction.
We count the distributed transaction ratios of workloads on the application side.
Figure~\ref{fig:tpcc-txlogic-dtx} shows the distributed transaction ratios of workloads generated by OLTP-Bench and Lauca.
Since OrderStatus and StockLevel are read-only transactions, the workloads used in Figure~\ref{fig:tpcc-txlogic-dtx} are only Payment transactions.
As can be seen from the results, when we turn off the parameter dependency information (Dep-Off), the distributed transaction ratio increases dramatically.
This is because the Warehouse ID parameters of four write operations in Payment transactions are randomly generated under Dep-Off, thus there is a high probability of becoming distributed transactions.
Overall, the control of transaction logic can effectively guarantee the similarity of distributed transaction ratio between synthetic workload and real workload.

% Overall, the control of transaction logic can ensure that the synthetic workload and the real workload have the same distributed transaction ratio in a distributed environment.

\iffalse
At present, there is no open source database system that can natively support distributed transactions and have a good support for TPC-C workloads.
For example, the performance of TiDB and Cockroach for TPC-C workloads is more than one order of magnitude lower than that of MySQL.
The lengthy execution process of TiDB and CokroachDB results in unacceptable latency of TPC-C transactions, so the distributed transaction ratio in such systems currently has little impact on system performance.
\fi

\subsection{Exploring Data Access Distribution} \label{sec:distexp}

\iffalse
\begin{figure}
	\centering
	\includegraphics[width=\columnwidth]{figures/ycsb-dist-throughput.eps}
	\caption{Dynamic workloads -- throughput}
	\label{fig:ycsb-dist-throughput}
\end{figure}

\begin{figure}
	\centering
	\includegraphics[width=\columnwidth]{figures/ycsb-dist-latency.eps}
	\caption{Dynamic workloads -- latency}
	\label{fig:ycsb-dist-latency}
\end{figure}

\begin{figure}
	\centering
	\includegraphics[width=\columnwidth]{figures/ycsb-dist-deadlock.eps}
	\caption{Dynamic workloads -- deadlock}
	\label{fig:ycsb-dist-deadlock}
\end{figure}
\fi

% In this paper, three types of data access distributions, namely S-Dist, D-Dist and C-Dist, are proposed to characterize the complex and various data access distributions of actual application workloads.
This section demonstrates the ability of proposed data access distributions, namely S-Dist, D-Dist and C-Dist, to depict the skewness, dynamics and continuity of data access.
Since the data access distribution of existing benchmark workloads is generally neither dynamic nor continuous, we build the evaluation workloads based on YCSB.
All experiments in this section are carried out on MySQL database with a test table from YCSB.
The size of test table is $10^6$, and the concurrency of database request is 20.

\begin{figure}
	\centering
	%\vspace{-1mm}
	\subfigure{
		\label{fig:ycsb-dist-throughput}
		\includegraphics[width=0.99\columnwidth]{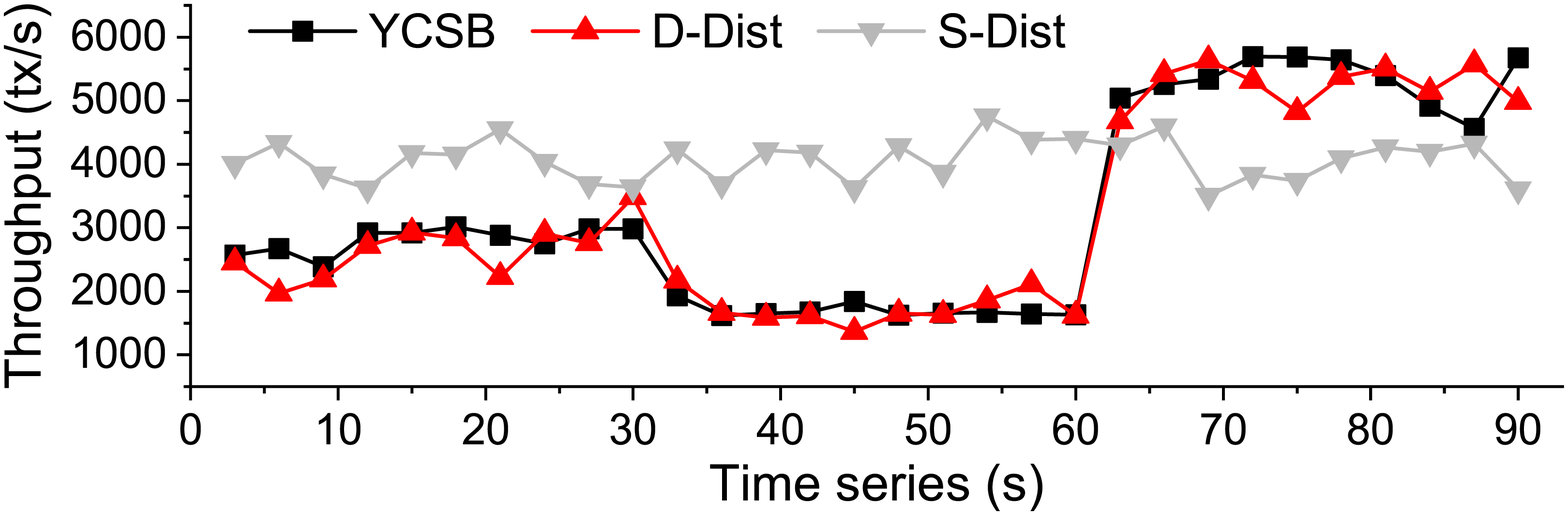}
		%\vspace{-5mm}
	}
	%\vspace{-1mm}
	\subfigure{
		\label{fig:ycsb-dist-latency}
		\includegraphics[width=0.99\columnwidth]{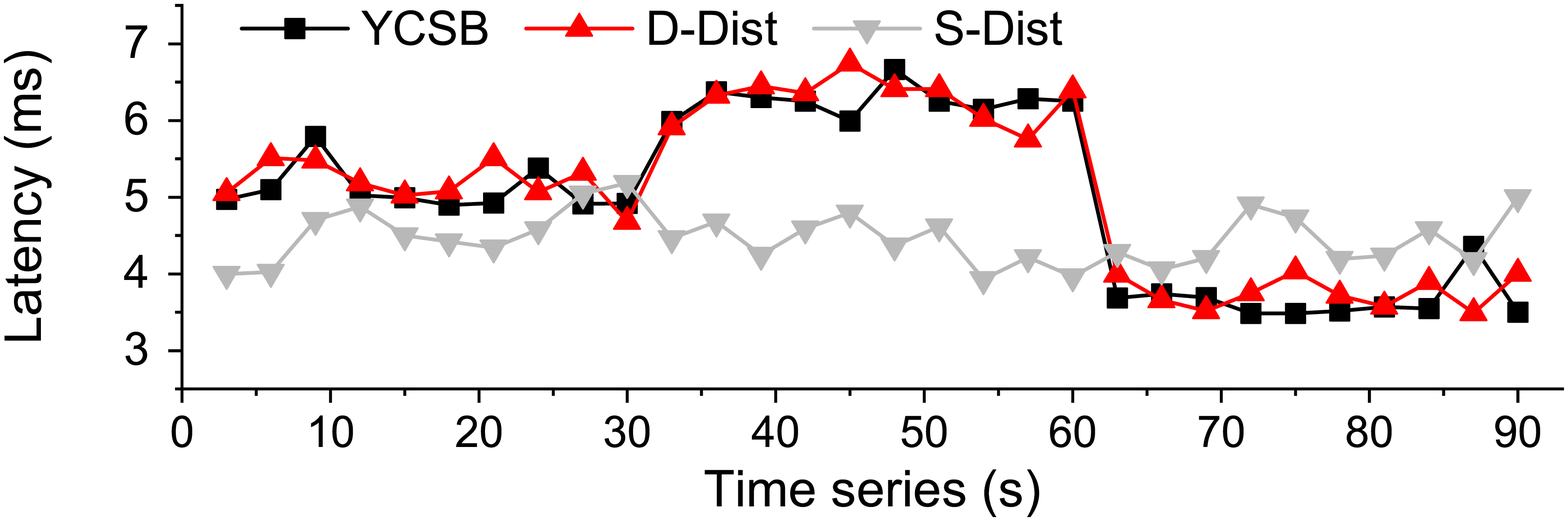}
		%\vspace{-5mm}
	}
	%\vspace{-1mm}
	\subfigure{
		\label{fig:ycsb-dist-deadlock}
		\includegraphics[width=0.99\columnwidth]{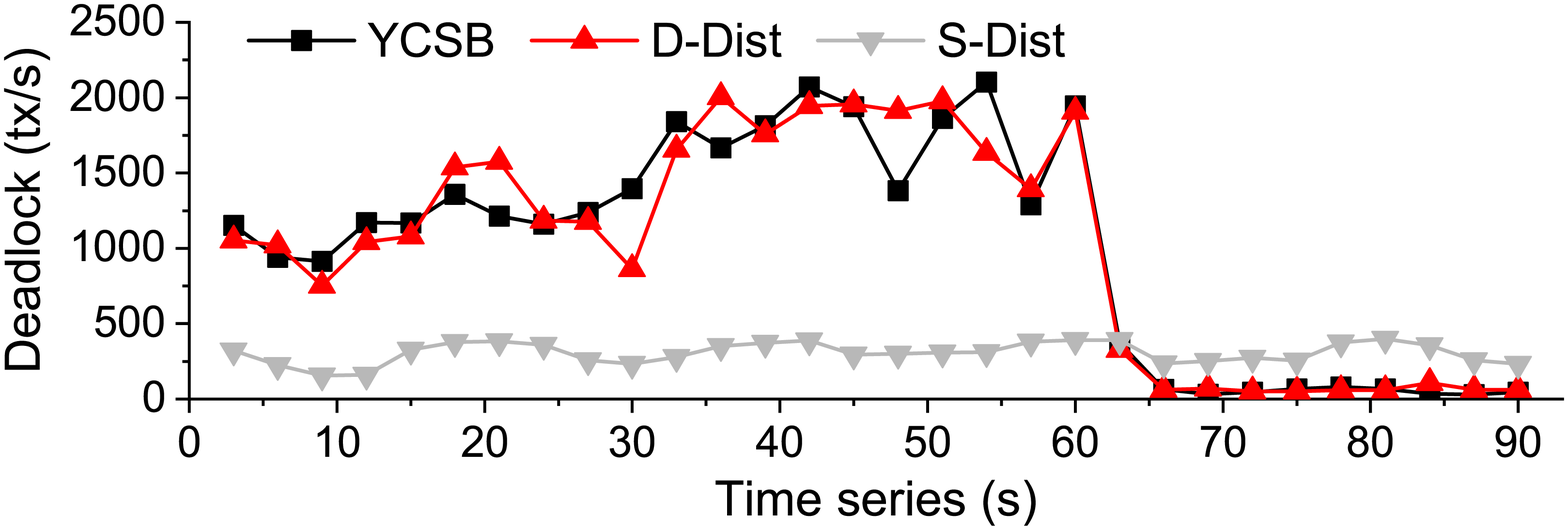}
	}
	%\vspace{-4.5mm}
	\caption{Exploring S-Dist and D-Dist for skewed and dynamic workloads}
	%\vspace{-3.5mm}
	\label{fig:ycsb-dist}
\end{figure}

The evaluation workload in Figure~\ref{fig:ycsb-dist} has only one type of transaction.
The transaction consists of five pairs of read-write operations, each of which reads a record first and then updates it.
% The size of test table is $10^6$, and the request concurrency is 20.
% Evaluation is carried out on the MySQL database.
The extended YCSB workload runs for 90 seconds, which is divided into three phases, and the data requests of each phase are upon $10^3$ records randomly selected.
In the first phase, the data access distribution is Zipf distribution with parameter $s$\;=\;$1$; the second phase is still the Zipf distribution, but the parameter $s$\;=\;$1.2$; and the third phase is the uniform distribution.
Figure~\ref{fig:ycsb-dist} shows the dynamic changes of transaction throughput, latency and deadlock throughput for workloads generated by YCSB and Lauca. % \ref{fig:ycsb-dist-throughput}-\ref{fig:ycsb-dist-deadlock} , respectively
Lauca has two groups of results, corresponding to S-Dist and D-Dist.
It can be seen from the results that when using D-Dist, the synthetic workload generated by Lauca is dynamically consistent with the real workload generated by YCSB on throughput, latency and deadlock, indicating that D-Dist can well depict the dynamics of workloads.
Meanwhile D-Dist is represented by S-Dist in each time window, which also shows that S-Dist can well characterize the skewness of workloads.
But the global S-Dist is not working well (grey lines in Figure~\ref{fig:ycsb-dist}), which is defined for the whole workload time and does not take into account the dynamic changes of the workload.

\begin{figure}
	\centering
	\includegraphics[width=0.66\columnwidth]{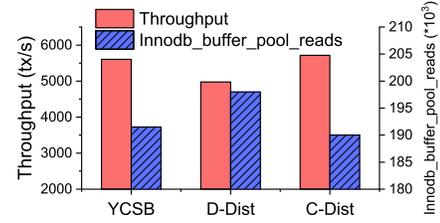}
	%\vspace{-3mm}
	\caption{Exploring C-Dist for continuous workloads}
	%\vspace{-2mm}
	\label{fig:ycsb-dist-cache}
\end{figure}

The evaluation workload in Figure~\ref{fig:ycsb-dist-cache} is the single-row update transaction of YCSB, by running 100 seconds with 1 second a time window.
The data requests in each time window is based on $10^3$ random records, and the selected records for each time window are 50\% coincident with the previous window.
The {\em Innodb\_buffer\_pool\_size} of MySQL is set to 16 MB.
Figure~\ref{fig:ycsb-dist-cache} presents the throughputs and {\em Innodb\_buffer\_pool\_reads} {\em increments} for workloads generated by YCSB and Lauca.
{\em Innodb\_buffer\_pool\_reads} is the number of logical reads that InnoDB cannot satisfy from the buffer pool, and have to read directly from disk.
From the results, we can see that the disk access of D-Dist is significantly higher than that of YCSB, and its throughput is lower.
This is because D-Dist is unable to catch the continuity of data access distribution, resulting in data requests in each time window are almost completely different with a low cache hit ratio.
The performance of C-Dist is consistent with YCSB, which indicates that C-Dist can well characterize the data access continuity.

\subsection{Performance of Lauca}

In this section, we use TPC-C workload traces to study the performance of Lauca, which is deployed on four servers.

%\vspace{-2mm}
\begin{figure}[H]
	\centering
	\begin{minipage}[t]{0.492\columnwidth}
		\includegraphics[width=\textwidth]{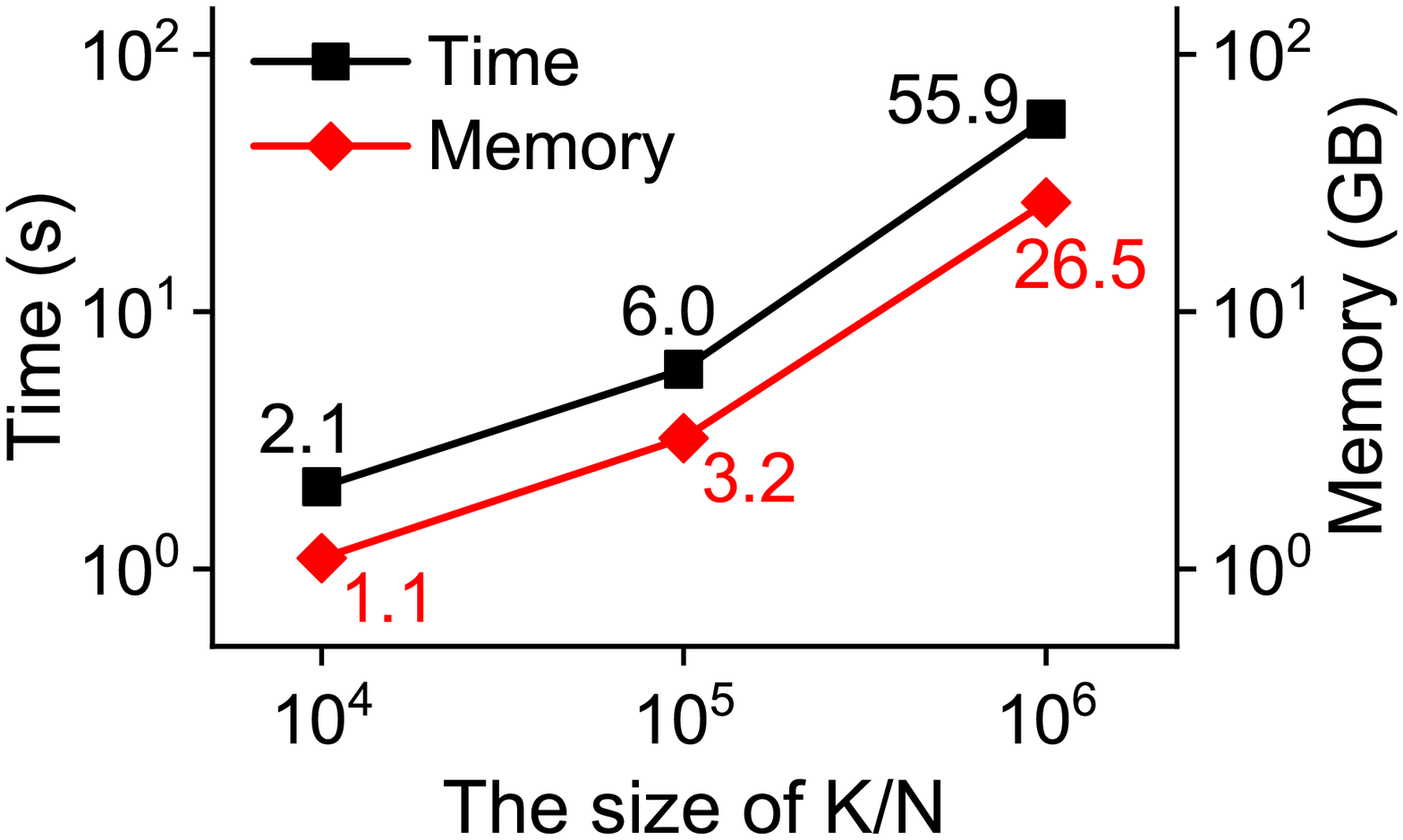}
		%\vspace{-6.5mm}
		\caption{\small Transaction logic\ \ extraction ($K$\;=\;$N$)}
		\label{fig:perf-logic}
	\end{minipage}
	\begin{minipage}[t]{0.492\columnwidth}
		\includegraphics[width=\textwidth]{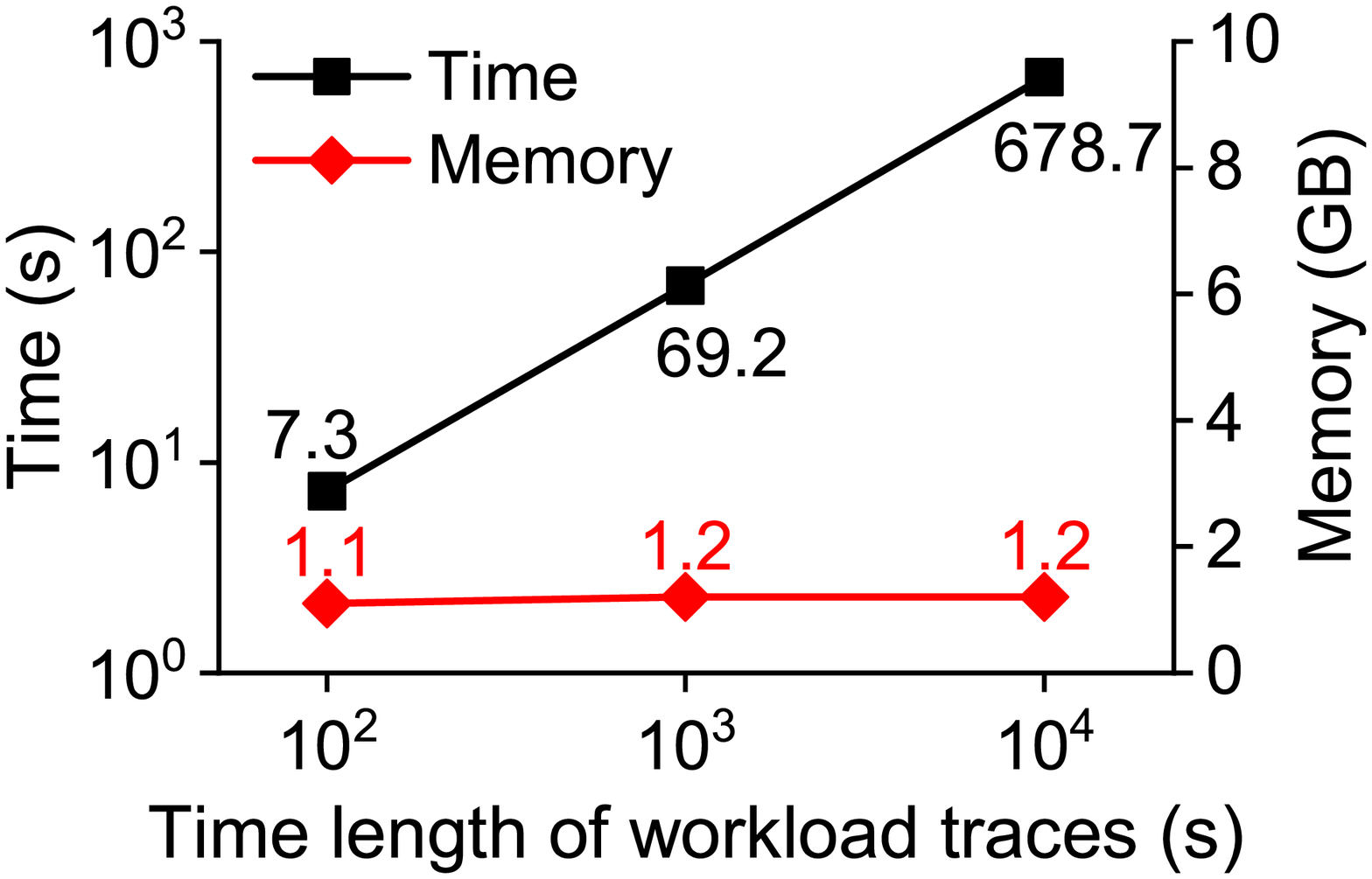}
		%\vspace{-6.5mm}
		\caption{\small Data access distribution C-Dist extraction}
		\label{fig:perf-dist}
	\end{minipage}
\end{figure}

%\vspace{-2mm}

Figure\ref{fig:perf-logic} shows the execution time and memory consumptions of transaction logic extraction under different sizes of $K$/$N$.
It can be seen from the results that when both $K$ and $N$ are $10^4$, the extraction time of transaction logic is only 2.1 seconds and the memory consumption is 1.1 GB.
With the increase of $K$ and $N$, the execution time and memory consumption increase almost linearly.
The experiments in Section~\ref{sec:workloadfidelity} are all conducted when $K$ and $N$ are set to $10^4$, and the high fidelity of generated workloads has been proved.
Overall, the transaction logic extraction in Lauca is efficient and can be done in a few seconds while guaranteeing the fidelity of synthetic workloads.

Figure\ref{fig:perf-dist} presents the execution time and memory consumptions of C-Dist extraction for workload traces with different time length. 
From the results, we can see that the extraction time of C-Dist is linear with the volume of workload traces while the memory consumption is constant.
This is because C-Dist is a window-based data access distribution, and the workload trace of a time window can be removed from memory after processing.
In Figure\ref{fig:perf-dist}, for TPC-C workload traces with the time length of $10^4$ seconds (transaction throughput is 3610.3 and log volume is 33.8 GB), the extraction time of C-Dist is 678.7 seconds and the memory consumption is 1.2 GB.
Since the maximum workload cycle in practical evaluations is generally one day, the result indicates that Lauca can effectively support the performance evaluation of high throughput workloads.

\vspace{0.5mm}
\section{Related Work} \label{sec:relatedwork}
\vspace{0.5mm}

There are a number of benchmarks for database performance evaluation in different application areas.
For OLAP applications, TPC-H, TPC-DS, and SSB~\cite{o2009star} are the frequently-used benchmarks with defined standard database schemas and test queries.
And there are TPC-C, TPC-E and SmallBank~\cite{alomari2008cost} benchmarks for evaluating the transaction processing capability of database systems. % ~\cite{b9tpcc}
In addition, CH-benCHmark~\cite{cole2011mixed} and HTAPBench~\cite{coelho2017htapbench} can provide a unified assessment for hybrid transaction/analytical processing (HTAP) systems.
% Along with the development of hybrid transaction/analytical processing (HTAP), CH-benCHmark~\cite{b15DBTest2011} and HTAPBench~\cite{b16ICPE2017} can provide a unified assessment for HTAP systems.
% Moreover, YCSB~\cite{b11SoCC2010} is usually used to measure the throughput of cloud service systems, whose workload is simple but requires high scalability.
% OLTP-Bench~\cite{b8VLDB2013} is a DBMS benchmarking testbed that provides implementations of popular OLTP benchmarks, making the performance evaluation of database more convenient.
However, the evaluation workloads of these standard benchmarks are abstractions for a class of applications, therefore they are too general to evaluate database performance for a specific application.
% {\color{blue}{A similar point about the discrepancy between current benchmarking practice and the real world workloads was made in a recent work by Vogelsgesang et al.~\cite{b27DBTest2018}.}}

In order to obtain elaborate workloads of a target application, workload trace replaying is an optional method.
Microsoft SQL Server equips two tools, i.e., SQL Server Profiler~\cite{profiler} and SQL Server Distributed Replay~\cite{replay}, for reproducing production workloads based on the SQL traces.
% , where SQL Server Profiler~\cite{profiler} is limited to replaying the workload from a single node but Distributed Replay~\cite{distributedreplay} can replay a workload from multiple nodes and better simulate a mission-critical workload.
Oracle Database Replay~\cite{galanis2008oracle} enables users to record workload traces on the production system with minimal performance impact and then to replay a full production workload that has the same concurrency and workload characteristics as the real one. % , b18DBTest2009
Due to data privacy issues, workload replay is difficult to apply in the practical database performance evaluation because it requires a real database state and original workload traces. % s~\cite{b3SIGMOD2008}
Additionally, workload extension (e.g., extending concurrency) is also a problem that is difficult to solve with current replay technologies.

Then workload simulation is necessary and urgent.
There are workload-aware data and query generators~\cite{binnig2007qagen, arasu2011data, lo2014mybenchmark, li2018touchstone} for database performance evaluation of OLAP applications. 
The input of these works generally includes database schema, basic data characteristics and size specifications for intermediate results of query trees.
The output is a synthetic database instance and instantiated test queries, conforming to the specified data and workload characteristics. % in the input
% And for OLAP applications, there are database scaling works~\cite{b21IS2013, b22PVLDB2016} which can scale up/down a given database instance, supporting application-specific database benchmarking.
Workload analyzers~\cite{yu1992workload, tran2015oracle} are designed to study and better understand the application workloads, but neither of which can generate synthetic workloads.
% Among which, REDWAR~\cite{yu1992workload} is developed to characterize the workload in a DB2 environment, and WI~\cite{tran2015oracle} is a tool for workload modeling and mining which is implemented in Oracle Database.
% However, neither of them~\cite{yu1992workload, tran2015oracle} generates synthetic workloads.
There are workload generators~\cite{jeong2005workload, ameri2016nowog} for database performance benchmarking.
% Jeong et al.~\cite{b24iiWAS2005} proposes a workload generator that helps database benchmarks be executed in a realistic environment by simulating hardware resource consumption status.
Jeong et al.~\cite{jeong2005workload} proposes a workload generator for simulating a realistic hardware resource consumption status.
% Rabl et al.~\cite{b25TPCTC2009} presents a benchmark tool that generates shifting eLearning application workloads to measure the adaptability of a database system.
NoWog~\cite{ameri2016nowog} introduces a workload description language for generating synthetic workloads benchmarking NoSQL databases.
None of these works~\cite{jeong2005workload, ameri2016nowog} can be used to simulate the various workloads of OLTP applications for application-oriented database performance evaluation.

\iffalse
There are some interesting non-relational workload generators~\cite{krishnamurthy2006synthetic, vogele2018wessbas, botta2012tool, bodik2010characterizing, gonccalves2016workload}.
For example, Krishnamurthy et al.~\cite{krishnamurthy2006synthetic} and WESSBAS~\cite{vogele2018wessbas} present automated approaches that aim to extract workload characteristics and construct synthetic workloads for session-based application systems.
Botta et al.~\cite{botta2012tool} provides a tool for generating realistic network workloads that can be used for the study of emerging networking scenarios.
In addition, SpikeGen~\cite{bodik2010characterizing} generates internet-scale workload spikes for evaluating the resiliency of stateful systems.
Moreover, CloudGen~\cite{gonccalves2016workload} can simulate both users' behavior and the network traffic of cloud storage services.
\fi

%\vspace{-0.5mm}
\section{Discussion} \label{sec:discussion}
%\vspace{-0.5mm}

% 参数依赖（事务逻辑）未考虑事务结构问题，比如依赖的数据项为空(相应的操作位于分支语句中)，此时用数据访问分布生成
% 数据访问分布未参数之间的关联关系
% 数据访问分布统计时未考虑事务逻辑的影响，应抛掉事务逻辑的影响来统计事务访问分布
% ......

\textbf{Data privacy protection.}
According to Lauca's workflow, only workload description information and workload statistics are exposed to testers during the evaluation. % (i.e., database schema and transaction templates)  (i.e., data characteristics, transaction logic and data access distribution)  basic workload description information
% Whether the data in the real database or the logged workload traces, it is untouchable for testers.
Neither the data in real databases nor the logged workload traces are visible to testers.
Components in Lauca that involve real application data are manipulated by data owners in the production environment.
Moreover, we can also anonymize the table names, column names and transaction names in both workload description information and workload statistics.
But at present, we cannot prevent doing reverse engineering on workload statistics to get sensitive information.

\textbf{Limitations of transaction logic.} %
The transaction logic in Definition~\ref{defn:txlogic} aims to support the common workloads in real world applications.
% But there are still three deficiencies in our current work.
% Below we list the deficiencies in our current work.
There are still three deficiencies in our current work.
Firstly, the relationships in Definition~\ref{defn:txlogic} are not complete.
For example, the relationship between two parameters may be represented by a quadratic function, which cannot be covered currently.
Secondly, only the relationship of two data items is considered at present, without considering the relationship among more data items. %, such as the sum of three parameters is a certain value.
Thirdly, the importance of different relationships may vary greatly and it is difficult to quantify them.
These issues are left to be addressed in future work.

\textbf{Choosing the window size and number of intervals.}
The setting of the time window size depends on how often the target workload changes.
If the workload changes frequently, the window size should be set to a small value, otherwise it can be set to a larger value.
% For instance, when the time window size is set to 1 second, we can capture the seconds-level workload change.
We recommend setting the time window size to 1 second so that even second-level workload changes can be captured. %, which is enough for most applications.
% and number of intervals
%Histogram is a commonly used statistical method for density estimation, which are widely used to represent data distribution statistics in industry databases, such as Oracle, MySQL and PostgreSQL. % ~\cite{b30Springer2013}
Histogram is widely used to represent data distribution statistics in industry databases, such as Oracle, MySQL and PostgreSQL.
%And we can apply academic and industrial mature methods~\cite{b30ESAIMPS2006, b31OracleHistograms} to choose the number of intervals.
At present, the number of intervals is selected by experiments.
We can apply existing academic and industrial methods~\cite{birge2006many, histograms} to choose the number of intervals in future work.

%\vspace{-0.5mm}
\section{Conclusion} \label{sec:conclusion}
%\vspace{-0.5mm}

In this paper, we presented Lauca, a transactional workload generator for application-oriented database performance evaluation.
Lauca uses transaction logic to depict the potential business logic of target applications, and data access distribution to characterize the access skewness, dynamics and continuity.
Our results on various workloads and popular databases show that Lauca consistently generates high-quality synthetic workloads.
% Future work focuses on improving the definition of transaction logic and proposing new types of data access distribution for covering more application workloads.

\bibliographystyle{abbrv}
\bibliography{sample-base}
\end{document}